\documentclass[apj]{emulateapj} 
\usepackage{graphicx}
\usepackage{apjfonts}
\usepackage{longtable}

\newcommand{\be}{\begin{equation}}
\newcommand{\ee}{\end{equation}}
\newcommand{\ba}{\begin{eqnarray}}
\newcommand{\ea}{\end{eqnarray}}

\newcommand{\kms}{\rm km\,s^{-1}}

\newcommand{\Mpc}{\rm Mpc}

\newcommand{\planck}{\textsl{Planck}}
\newcommand{\wmap}{\textsl{WMAP}}
\newcommand{\lcdm}{$\Lambda$CDM}

\def\ls{\mathrel{\hbox{\rlap{\hbox{\lower4pt\hbox{$\sim$}}}\hbox{$<$}}}}
\def\gs{\mathrel{\hbox{\rlap{\hbox{\lower4pt\hbox{$\sim$}}}\hbox{$>$}}}}
\shorttitle{\wmap\ Observations of \planck\ ESZ Clusters}
\shortauthors{Ma, Hinshaw, \& Scott}
\begin{document}
\title{\wmap\ Observations of \planck\ ESZ Clusters}

\author{
 Yin-Zhe Ma,$\!$\altaffilmark{1,2}
  Gary Hinshaw,$\!$\altaffilmark{1}
  Douglas Scott,$\!$\altaffilmark{1}
}

\altaffiltext{1}{Department of Physics and Astronomy, University
of British Columbia, Vancouver, BC, V6T 1Z1, Canada}
\altaffiltext{2}{Canadian Institute for Theoretical Astrophysics,
Toronto, Canada}

\begin{abstract}
We examine the Sunyaev--Zeldovich (SZ) effect in the 7-yr
\textit{Wilkinson Microwave Anisotropy Probe} (\wmap\ ) data by
cross-correlating it with the \planck\ Early-release
Sunyaev--Zeldovich catalog. Our analysis proceeds in two parts. We
first perform a stacking analysis in which the filtered \wmap\
data are averaged at the locations of the 175 \planck\ clusters.
We then perform a regression analysis to compare the mean
amplitude of the SZ signal, $Y_{500}$, in the \wmap\ data to the
corresponding amplitude in the \planck\ data. The aggregate
\planck\ clusters are detected in the 7-yr \wmap\ data with a
signal-to-noise ratio of 16.3.  In the regression analysis we find
that the SZ amplitude measurements agree to better than 25\%:
$a=1.23 \pm 0.18$ for the fit $Y^{\rm{wmap}}_{500} =
aY^{\rm{planck}}_{500}$.
\end{abstract}


\section{Introduction}
\label{intro}
Clusters of galaxies are important objects for studying cosmology
and large-scale structure formation. In hot clusters, around
$12$\% of their mass is in the form of a hot, ionized
intra-cluster medium (ICM, \citealt{McCarthy07}). The ICM can be
studied using direct X-ray imaging and/or the Sunyaev-Zeldovich
effect (SZ, \citealt{Sunyaev72,Sunyaev80}). Observations of the
latter have grown tremendously in recent years thanks to an array
of powerful new instruments
\citep{Birkinshaw78,Birkinshaw99,Carlstrom02}.

The thermal SZ effect is a secondary anisotropy in the cosmic
microwave background (CMB) radiation, caused by CMB photons
scattering off free electrons in the hot ICM through the inverse
Compton effect.  The effect boosts the photon's energy and thus
distorts the CMB spectrum in the direction of a cluster, causing a
temperature decrement on the low frequency side of the CMB peak
(specifically, $\nu < 217$ GHz), and an increment at high
frequencies.


The SZ effect is especially powerful for studying high-redshift
galaxy clusters.  Since the Compton $y$-parameter (the integral of
the ICM pressure along the line-of-sight) does not diminish with
distance, and since the CMB has a nearly uniform surface
brightness, the SZ effect does not diminish with increasing
redshift.  This makes SZ surveys especially suitable for finding
high redshift clusters \citep{Planck11a}.  These cluster samples
can be used to constrain cosmological models (e.g., from the
evolution of the mass function) and to probe the physics of
structure formation (e.g., from cluster scaling relations and
structural properties, \citealt{Planck11a}).  Many ongoing
experiments are measuring the SZ effect; for instance, the South
Pole Telescope (SPT, \citealt{Carlstrom11}) and the Atacama
Cosmology Telescope (ACT, \citealt{Marriage11}) are each measuring
tens to hundreds of clusters over a few hundred square degrees
\citep{Haiman01,Weller02,Levine02,Majumdar04,Douspis06,Shaw11}.

With the mission to precisely measure CMB temperature and
polarization anisotropy, the \planck\ satellite was successfully
launched by the European Space Agency (ESA) on 14 May 2009
\citep{Planck05}.  \planck\ carries a scientific payload
consisting of 74 detectors sensitive to frequencies between 25 and
1000 GHz. \planck\ scans the sky continuously with an angular
resolution between about $30$ arcmin (FWHM) at the lowest
frequency to about $4$ arcmin at the highest \citep{Planck11a}.
Its combination of frequency coverage, sensitivity, and angular
resolution, enables it to measure the spatial anisotropy of the
CMB with an accuracy set by fundamental astrophysical limits.

Performing an SZ cluster survey over the full sky was an important
goal for \planck\ \citep{Aghanim97}, and the Project produced its
Early Release Compact Source Catalog (ERCSC, \citealt{Planck11b})
in 2011.  The ERCSC included a catalog of 189 SZ clusters detected
with high-reliability from the first ten months of the \planck\
survey \footnote{Planck ERCSC website:
http://www.rssd.esa.int/Planck}.  The Project is ultimately
expected to release a few thousand high-reliability SZ clusters
\citep{Planck11d}.

The \wmap\ mission was launched by NASA on 30 June 2001 to map the
CMB anisotropy over the full sky to multipole moment $\ell \sim
1000$ (angular resolution $\sim 0.2^{\circ}$). \wmap\ has
precisely measured the cosmological parameters to unprecedented
accuracy, but its angular resolution and frequency coverage ($23$
to $94$ GHz) were not optimized for SZ detection.  \wmap\
independently detected a dozen SZ clusters, all of which were
well-known \citep{Komatsu11}.  It is therefore interesting and
important to see if the SZ clusters detected by \planck\ are also
detected in the 7-yr \wmap\ data.  This is the aim of this paper.


The plan of this paper is as follows. In Section \ref{SZcluster},
we review the main features of the SZ effect and discuss the
universal pressure profile we use to model the ICM pressure. In
Section \ref{datadescrib}, we review the \planck\ ESZ catalog and
the \wmap\ W-band data which form the basis of our analysis. In
Section \ref{filter}, we present the matched filter we apply to
the \wmap\ sky map and examine the filtered model profiles for a
range of cluster parameters. We present our main results  in
Section~\ref{result}, and some concluding remarks follow.

Throughout this paper, we adopt a fiducial flat $\Lambda$CDM
cosmology with Hubble constant $H_0=70$  $\kms \Mpc^{-1}$, and
matter density parameter $\Omega_{\rm{m}}=0.3$. The Hubble
parameter at redshift $z$ is $H(z) = H_0 E(z)$, where $E^{2}(z) =
\Omega_{\rm{m}}(1+z)^3 + \Omega_{\Lambda}$.

\section{Cluster pressure profile}
\label{SZcluster}

\subsection{SZ effect}
\label{SZeffect}

The thermal SZ effect is a secondary anisotropy in the the cosmic
microwave background (CMB) radiation, caused by CMB photons
scattering off free electrons in the hot ICM through the inverse
Compton effect.  The effect boosts the photon's energy and thus
distorts the CMB spectrum, causing a temperature decrement on the
low frequency side of the CMB peak, and vice versa.  The induced
temperature anisotropy in the direction of a cluster is
\citep{Sunyaev72,Birkinshaw99,Carlstrom02} \be \frac{\Delta T}{T}
= \left[\eta \frac{e^{\eta}+1}{e^{\eta}-1} - 4\right] y \equiv
g_{\nu}y, \label{deltaT1} \ee where $g_{\nu} \equiv
[\eta(e^{\eta}+1)/(e^{\eta}-1)] - 4$ captures the frequency
dependence, and \be \eta = \frac{h \nu}{k_{\rm{B}}T_{\rm{CMB}}} =
1.76 \left(\frac{\nu}{100 \textrm{ GHz}}\right). \label{xdefine}
\ee The dimensionless Comptonization parameter $y$ depends on the
electron temperature, $T_{\rm{e}}({\bf r})$, and density,
$n_{\rm{e}}({\bf r})$, in the ICM, \be y = \int n_{\rm{e}}({\bf
r})\sigma_{\rm{T}} \,\frac{k_{\rm{B}} T_{\rm{e}}({\bf
r})}{m_{\rm{e}} c^2} \,dl, \label{comptony} \ee where
$\sigma_{\rm{T}}$ is the Thomson cross section, $k_{\rm{B}}$ is
the Boltzmann constant, $m_{\rm{e}}$ is the electron rest mass and
the integral is taken along the line of sight. Since $y$ is
positive, the sign of $g_{\nu}$ determines whether $\Delta T$ is
an increment or decrement in the CMB temperature.  With $T_{\rm
CMB} = 2.725$ K, we have $g_{\nu}>0$ for $\nu > 217$ GHz, and vice
versa, and $g_{\nu} \rightarrow -2$ at low frequencies ($\eta \ll
1$).



It is convenient to define the integrated SZ signal as $Y \equiv
\int y\,d\Omega$, where the integration is over the solid angle of
the cluster.  This is equivalent to a volume integral \be Y =
\frac{1}{D_{\rm{A}}^2}\frac{\sigma_{\rm{T}}}{m_{\rm{e}} c^2} \int
P dV, \ee where $D_{\rm{A}}$ is the angular diameter distance to
the system and $P = n_{\rm{e}} k T_{\rm{e}}$ the electron pressure
\citep{Planck11a}. In the following, we denote the integral
performed over the sphere of radius $R_{500}$\footnote{$R_{500}$
is defined as the radius within which the density contrast is
$>500$.} ($5R_{500}$), as $Y_{500}$ ($Y_{5R500}$).  Note that $Y$
has units of solid angle, typically arcmin$^2$.

\subsection{Universal profile}
\label{Univerpro}

\begin{figure*}
\centerline{
\includegraphics[bb=0 0 509 337, width=3.2in]{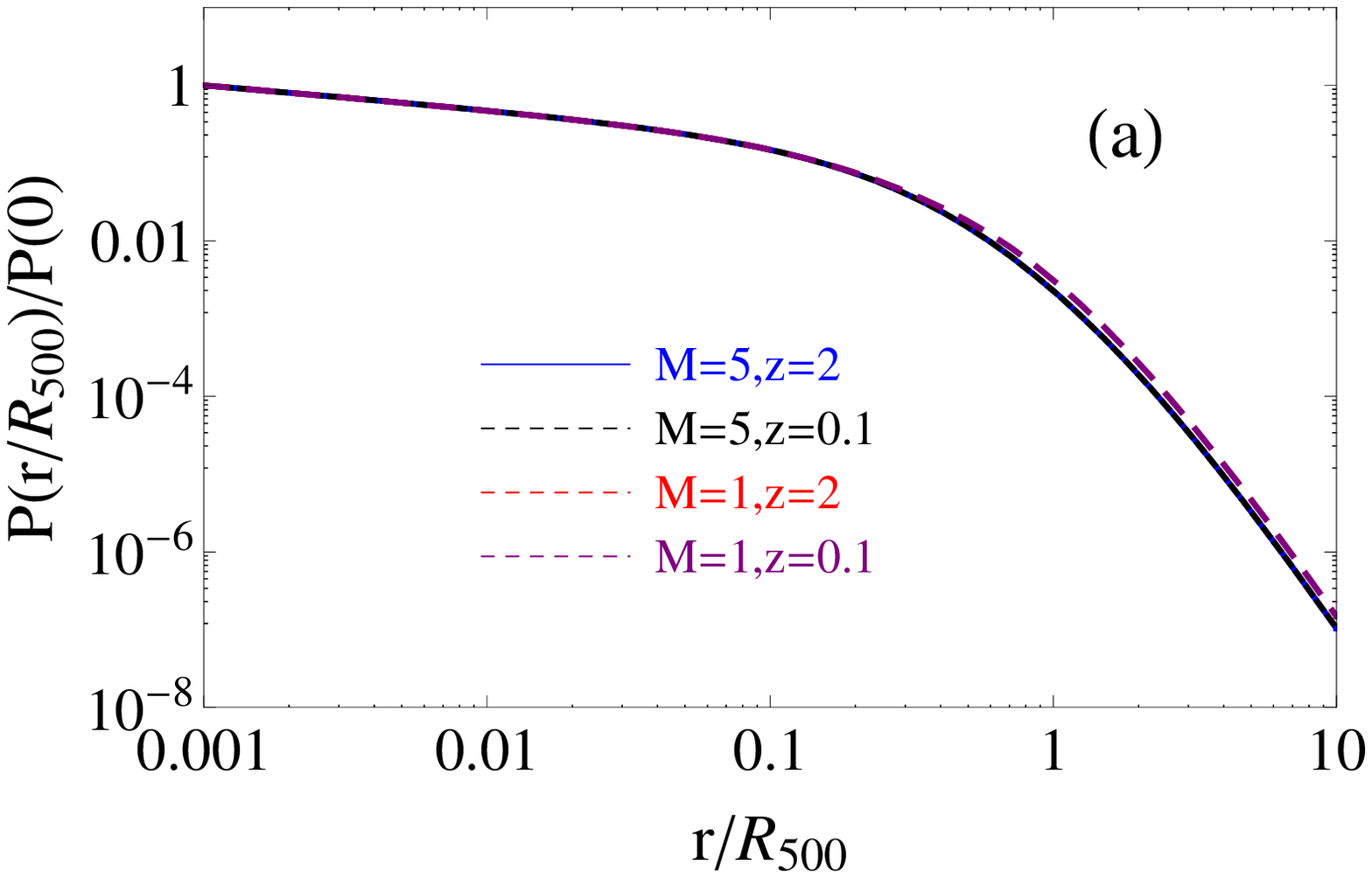}
\includegraphics[bb=0 0 509 337, width=3.2in]{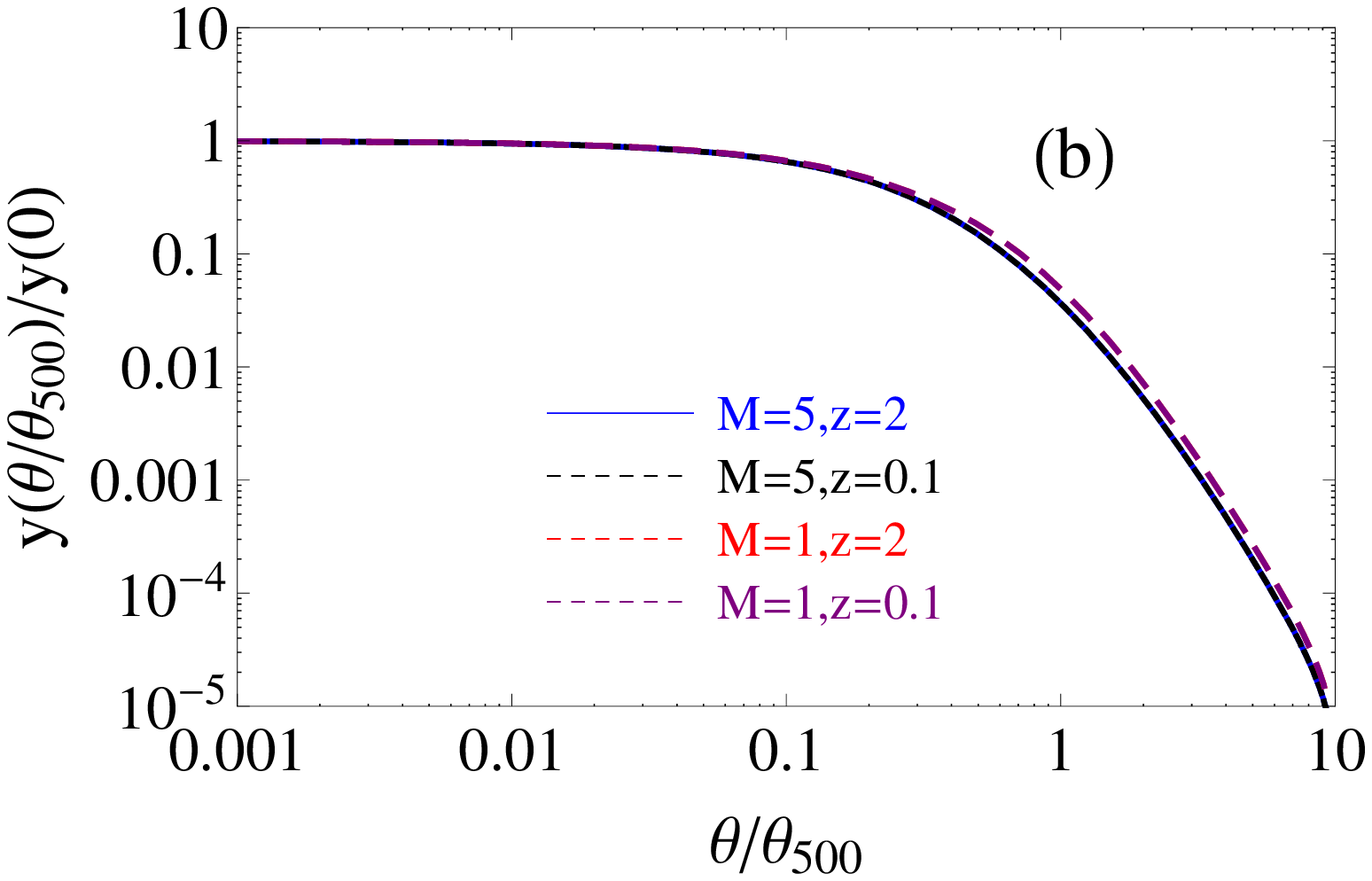}}
\caption{{\it Left} -- Pressure profile from Eq.~(\ref{unipres})
for four hypothetical clusters. {\it Right} -- The temperature
profile from Eq.~(\ref{ypro}) for the same four cluster
parameters. The units of $M$, $R_{500}$, and $\theta_{500}$ are $3
\times 10^{14} h^{-1}_{70} \rm{M}_{\odot}$, Mpc, and arcmin,
respectively, where $h_{70}=(h/0.7)$. \label{fig:press}}
\end{figure*}

In order to study the SZ effect in the \wmap\ data, we need a
model for the pressure profile of the cluster gas, $P({\bf r})$.
\cite{Arnaud10} studied ICM pressure profiles using a sample of 33
local ($z < 0.2$) clusters observed with \textit{XMM-Newton}.  The
sample spans a mass range of $10^{14} \rm{M}_{\odot} <
\textit{M}_{500} < 10^{15} \rm{M}_{\odot}$, where $M_{500}$ is the
mass enclosed within $R_{500}$ (assumed to be spherically
symmetric).  The pressure profiles in their sample can be
described by a universal profile that is scaled with both mass-
and redshift-dependent factors. The dispersion of the data about
the scaled profile is less than 30\% beyond $0.2 R_{500}$
\citep{Arnaud10}.

With $x\equiv r/R_{500}$, the form of the universal profile given by \citet{Arnaud10} is
\ba
P(x) = && 1.65 \times 10^{-3} E(z)^{\frac{8}{3}} \left[\frac{M_{500}}{3\times 10^{14}M_{\odot }h_{70}^{-1}}\right]^{\frac{2}{3}+\alpha_{\rm{p}}+\alpha _{\rm{p}}^{\prime}} \nonumber \\
& & \times \, \tilde{p}(x) \, h_{70}^{2}\text{ }\left[\textrm{keV
cm}^{-3}\right], \label{unipres} \ea where $h_{70}=(h/0.7)$,
$\alpha_{\rm{p}} =0.12$, and \be \alpha _{\rm{p}}^{\prime} = 0.1 -
(\alpha_{\rm{p}}+0.1) \, \frac{(x/0.5)^3}{1+(x/0.5)^3}. \ee Here
$\tilde{p}(x)$ is the generalized NFW model proposed by
\citet{Nagai07} (see also \citealt{Arnaud10}) \be \tilde{p}(x) =
\frac{P_0}{(c_{500} x)^{\gamma}\left[1+(c_{500}
x)^{\alpha}\right]^{(\beta-\gamma)/\alpha}}, \label{px} \ee where
$P_{0} = 8.403 h_{70}^{-3/2}$ is the overall magnitude of the
pressure profile, and $c_{500}, \gamma, \alpha$, and $\beta$
determine the slope of the profile. By fitting this pressure
template to the simulated profile, \cite{Arnaud10} found the best
parameters are $c_{500} = 1.177,$ $\gamma = 0.3081,\alpha =
1.051$, and $\beta = 5.4905$.

The only parameter left undetermined in Eq.~(\ref{unipres}) is the
mass parameter $M_{500}$, which is nearly degenerate with the
overall pressure normalization term $P_0$ in Eq.~(\ref{px}).  To
illustrate this degeneracy, we plot in Fig.~\ref{fig:press}a the
normalized pressure profiles, $P(x)/P(0)$, for several
combinations of $M_{500}$ and cluster redshift, $z$.  They show
very similar normalized profiles.

%

%

The angular profile of the model SZ signal is obtained by
projecting the 3-D pressure profile $P({\bf r})$ onto the plane of
the sky and calculating the corresponding temperature profile.
Following \citet{Komatsu11}, the projected profile, in keV
cm$^{-3}$, is \ba P_{\rm 2d}(\theta) = \int_{-\sqrt{r_{\rm
out}^2-\theta^2 D_{\rm{A}}^2}}^{\sqrt{r_{\rm out}^2-\theta^2
D_{\rm{A}}^2}} P\left(\sqrt{l^2 + \theta^2 D_{\rm{A}}^2}\right) \,
dl, \label{pe} \ea where $D_{\rm{A}}$ is the angular diameter
distance to redshift $z$, and $r_{\rm out}$ is the truncated
radius, which we take to be $r_{\rm out} = 6R_{500}$.  Beyond this
range, the projection is not very sensitive to $r_{\rm out}$
because the profile falls off rapidly.

Given the 2-dimensional pressure profile, the temperature profile
is \be \frac{\Delta T_{\rm{SZ}}(\theta)}{T} = g_{\nu} \,
\frac{\sigma_{\rm{T}}}{m_{\rm{e}}c^2} \, P_{\rm{2d}}(\theta),
\label{tempro} \ee and the Comptonization parameter $y$
(Eq.~(\ref{comptony})) is \ba
y(\theta) & = & \frac{\sigma_{\rm{T}}}{m_{\rm{e}}c^2} \, P_{\rm{2d}}(\theta) \nonumber \\
& = & \left(8 \times 10^{-3}\right) \int_0^{\sqrt{(6\theta_{500})^2 - \theta^2}\tilde{D}_{\rm{A}}} dx \nonumber \\
& \times &
P\left(\sqrt{\left(\frac{x}{\theta_{500}\tilde{D}_{\rm{A}}}\right)^2
+ \left(\frac{\theta}{\theta_{500}}\right)^2}\right), \label{ypro}
\ea where we have used $\theta \equiv r/D_{\rm{A}}$.

In Fig.~\ref{fig:press}b, we plot the normalized Comptonization
profile, $y(\theta)/y(0)$ vs. $\theta/\theta_{500}$, for the same
four cluster parameters as in Fig.~\ref{fig:press}a. Again, the
normalized $y$ profiles are very similar.  Since we fit the
normalization of the profile to the \wmap\ data, we are primarily
concerned with the angular radius of each cluster, $\theta_{500}$,
which scales the extent of the profile.  As discussed below, this
parameter may be derived from information provided in the
\textit{Planck} ESZ catalog.  Thus, when fitting the \wmap\ data,
we adopt the normalized profile in Eq.~(\ref{ypro}), using the
$\theta_{500}$ value predicted by \planck.



\section{Data description}
\label{datadescrib}

\begin{figure*}
\centerline{
\includegraphics[bb=0 0 606 400, width=3.2in]{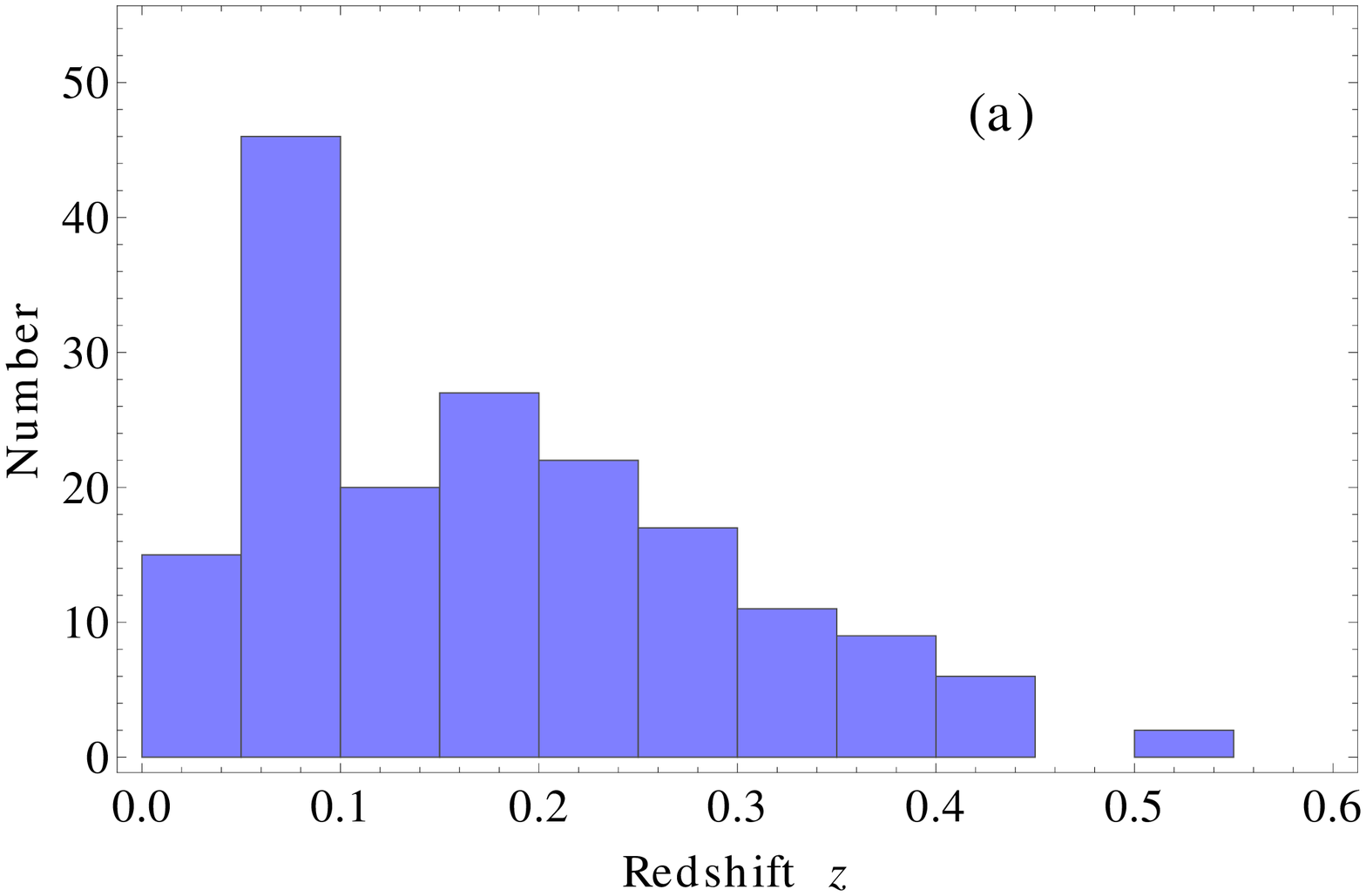}
\includegraphics[bb=0 -50 525 280, width=3.2in]{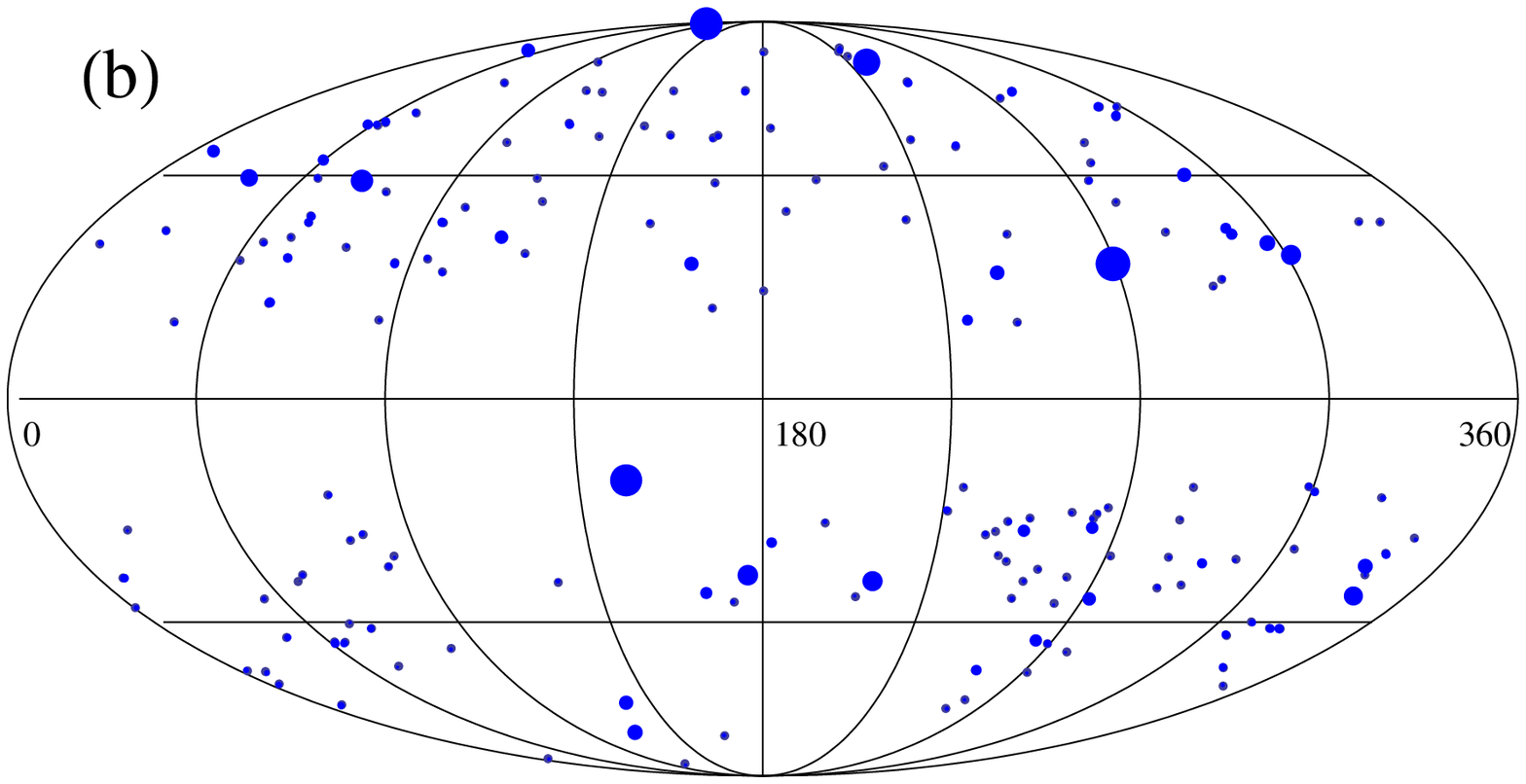}}
\caption{{\it Left} -- redshift histogram of $175$ \planck\ ESZ
clusters. {\it Right} -- full-sky distribution of \planck\ ESZ
clusters; the size of each point is proportional to the angular
radius $\theta_{500}$. \label{fig:szcluster}}
\end{figure*}

In this section, we introduce the data used in our analysis: the
\planck\ Early SZ catalog, which consists of $189$ clusters; and
the \wmap\ W-band sky map that will be used for consistency
testing.

\subsection{\planck\ ESZ catalog}
\label{planckdata}

\planck\ is a full-sky CMB survey with nine different frequency
channels, 30, 44, 70, 100, 143, 217, 353, 545, and 857 GHz.  The
FWHM angular resolution for these channels are 33, 24, 14, 10,
7.1, 5, 5 and 5 arcmin, respectively \citep{Planck05}.

The ESZ catalog was obtained from a blind, multi-frequency search
of the \planck High Frequency Instrument (HFI) maps, using an
all-sky extension of the algorithm given in \citet{Melin06}.  The
team used a matched multi-frequency filter method to enhance
signal-to-noise in the $189$ detected objects; the filter
optimizes detectability by using a linear combination of frequency
maps to null the CMB signal, and spatial filtering to suppress
foregrounds and instrument noise \citep{Planck11a}. Therefore, the
\planck\ filter boosts the expected SZ signal over all-sky
emission (including foregrounds) and noise. In principle, SZ
clusters can be also seen in the three Low Frequency Instrument
(LFI) channels; however, the low-frequency beam size ($\sim$25
arcmin) is generally much larger than a typical cluster size of
$\sim$5 arcmin, effectively diluting the signal, and therefore the
LFI channels were not included in the analysis \citep{Planck11a}.
The position and radius ($5R_{500}$) of each cluster profile was
varied to maximize the signal-to-noise ratio of each detection.
The position, angular radius, $5\theta_{500}$, and integrated
Comptonization parameter, $Y_{5R500}$, are tabulated for each
cluster in the catalog \citep{Planck11a}\footnote{With the
assumption of spherical symmetry in the model profile, we have
$Y_{5R500} = 1.8 \times Y_{500}$.}.

The preliminary analysis yielded 201 high signal-to-noise
($S/N>6$) candidates, of which 189 were deemed to be of high
reliability.  Of these, 169 were already known from X-ray and
optical surveys, while 20 clusters were newly detected.  The new
clusters were subsequently confirmed by \textit{XMM-Newton}
observations \citep{Planck11c} and by the Arcminute Microkelvin
Imager (AMI) survey \citep{Zwart08}.

Redshifts are known for 175 of the 189 clusters.  Since we require
redshift information to construct a pressure profile, we limit our
analysis of \wmap\ data to this subset of the full catalog.  The
redshift distribution and positions of the \planck\ ESZ clusters
are shown in Fig.~\ref{fig:szcluster}: in  panel (a), the redshift
distribution of these clusters is seen to range from $0.01$ to
$0.5$ with a mean redshift of 0.18.  Their spatial distribution
shown in panel (b) is close to uniform across the sky, outside a
$|b|<14^{\circ}$ Galaxy cut, although in principle more objects
are sampled around the ecliptic poles.

\subsection{\wmap\ W-band data}
\label{wmapwband}

\begin{figure*}
\centerline{
\includegraphics[bb=0 0 744 438, width=3.2in]{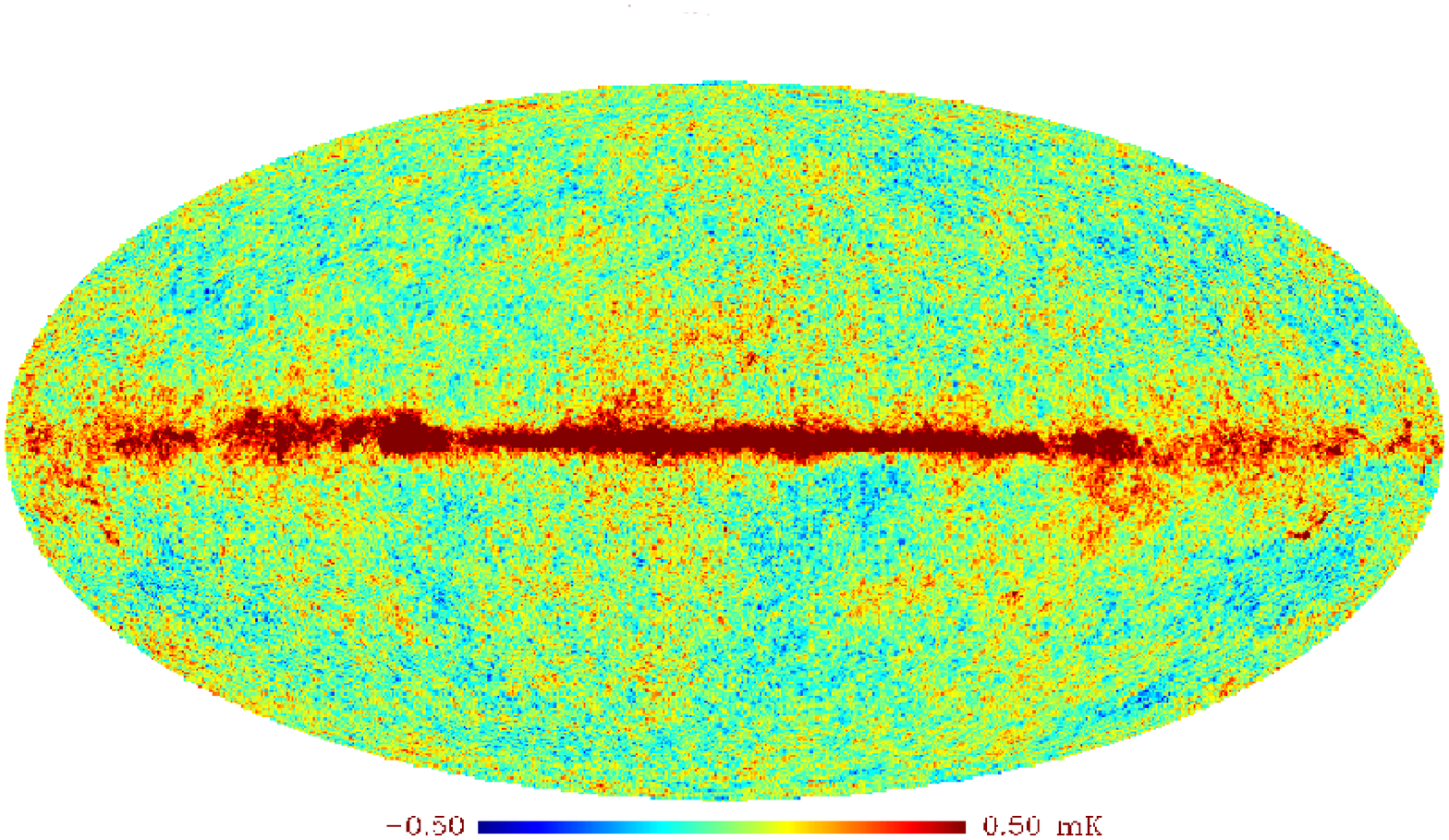}
\includegraphics[bb=0 0 744 438, width=3.2in]{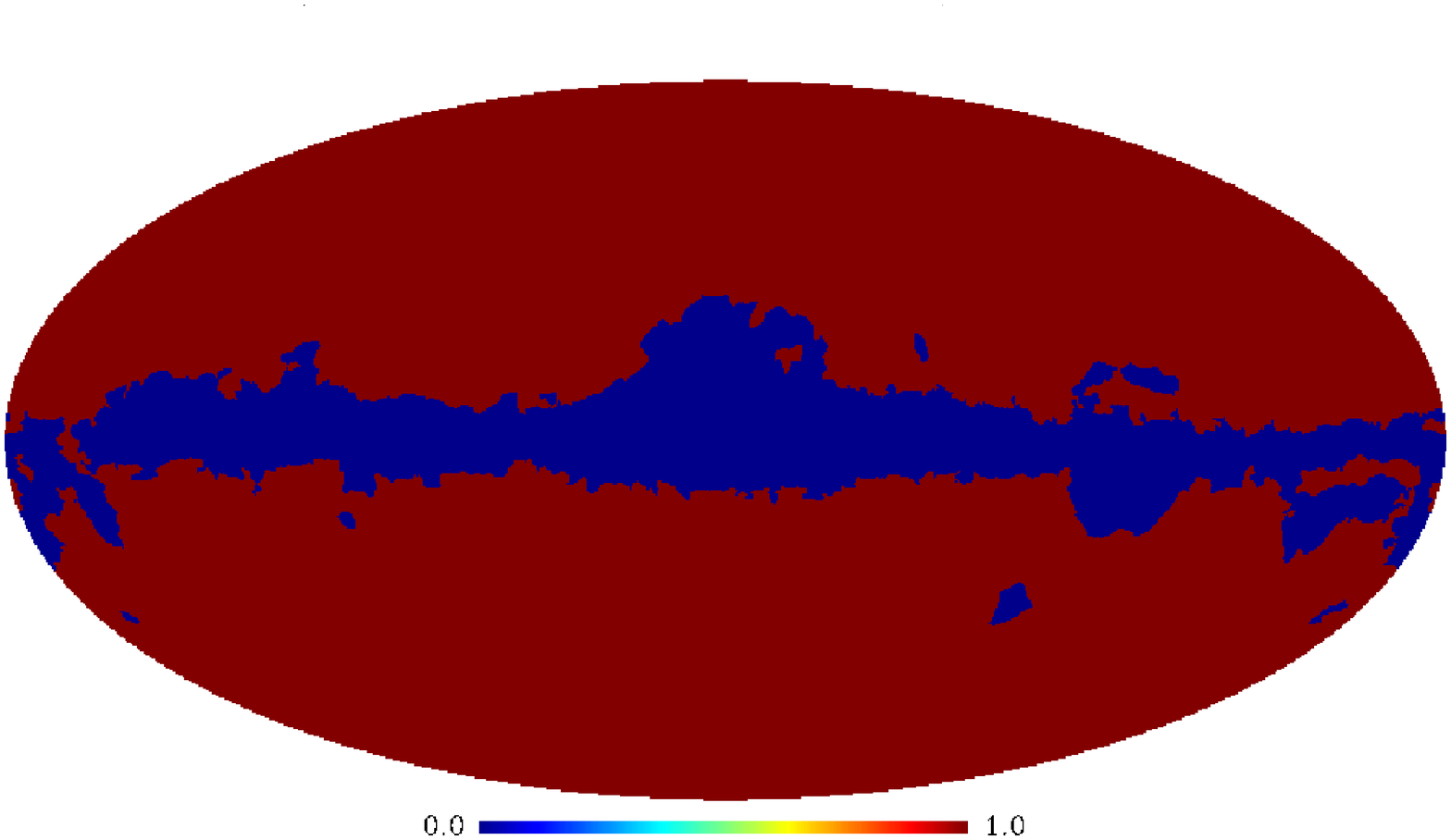}}
\centerline{
\includegraphics[bb=0 0 744 438, width=3.2in]{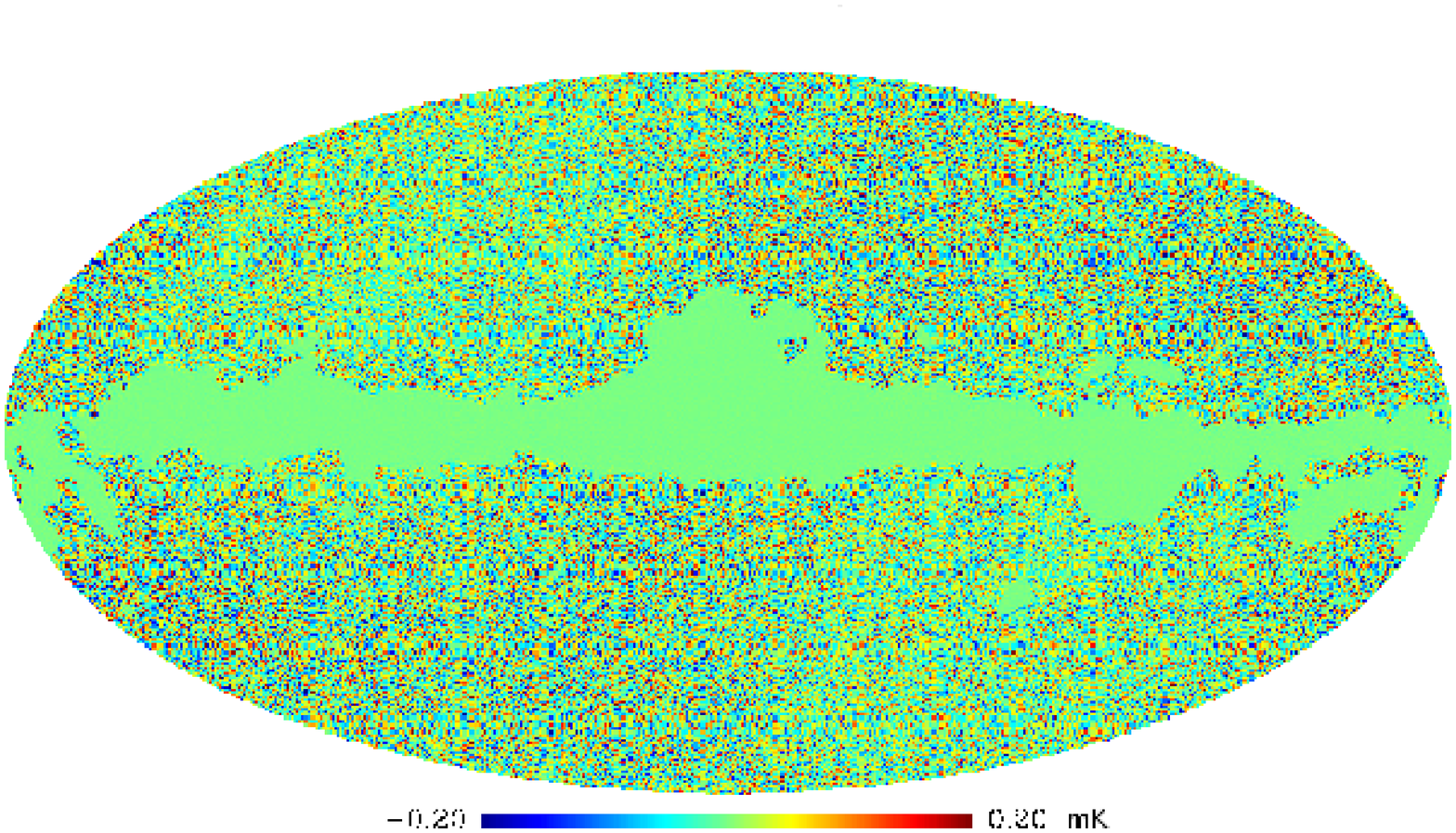}
\includegraphics[bb=0 0 744 438, width=3.2in]{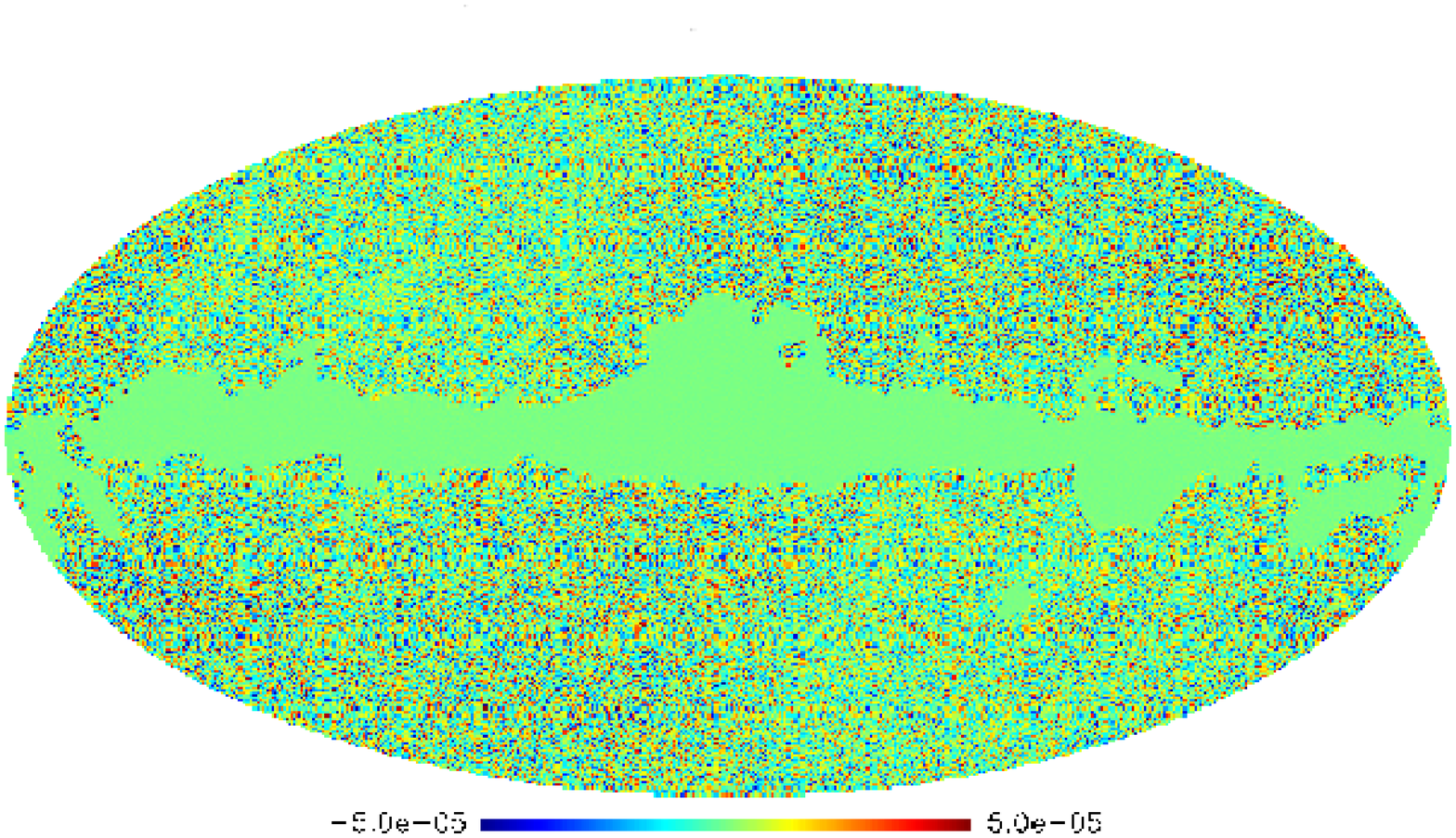}}
\caption{\wmap\ data used in this analysis.  {\it Upper-left} --
co-added 7-yr W-band map. {\it Upper-right} -- \wmap\ 7-yr
analysis mask, including point source cuts, 82.4\% of the sky
remains. {\it Lower-left} -- filtered W-band temperature map, as
per Eq.~(\ref{deltaT3}). {\it Lower-right} -- dimensionless
$y$-parameter map, as per Eq.~(\ref{ypro}).} \label{fig:maps}
\end{figure*}

The \wmap\ satellite produced full-sky maps at five frequency
bands from 23 to 94 GHz, with a FWHM angular resolution from 52.8
to 13.2 arcmin, respectively. These maps were used to measure
cosmological parameters with unprecedented accuracy.

Taking into account the combined effects of spectral shape,
$g_{\nu}$, and diffraction-limited beam size, the largest SZ {\em
decrement} occurs around 140 GHz \citep{Carlstrom02}, which is
higher than the highest \wmap\ frequency.  Within the \wmap\
bands, the 94 GHz W-band map has the most sensitivity to SZ
signal, owing mostly to its angular resolution (and relatively low
foreground contamination). We use the W-band {\tt HEALPix} map at
$n_{\rm{side}}=1024$ \citep{Gorski05}. In the following analysis,
we fit SZ profiles to the 7-yr W-band sky map.  At 94 GHz, the SZ
decrement is $\Delta T_{\rm{SZ}}/T = -1.56 \, y$.

\section{Filtering technique}
\label{filter}

To optimally characterize the SZ signal, we need to filter the
observed maps, which are dominated by primary CMB fluctuations,
but which also include residual foreground signals (including
extragalactic point sources) and instrument noise.  In this
section we describe the choice of filter (following
\citealt{Tegmark98}), calculate the temperature profile of SZ
clusters after filtering, and compute the $Y$-parameter.  We then
compare $Y_{500}$ derived from \wmap\ with that derived from
\planck.

\subsection{Optimal filter}
\label{matchfil}

\begin{figure}
\centerline{\includegraphics[bb=0 0 545
353,width=3.2in]{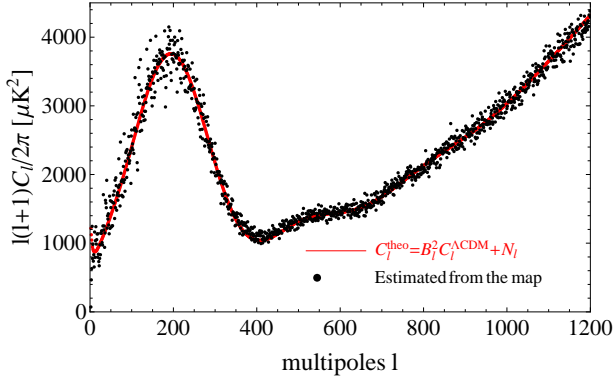}} \caption{Measured (black) and
predicted (red) power spectra, $C_{\ell}$, from the \wmap\ 7-yr
W-band map.  The predicted spectrum is based on the best-fit
\lcdm\ model convolved with the beam, plus detector noise,
assuming $N_{\ell} \equiv 0.0187$ $\mu$K$^2$ \citep{Hinshaw07}.}
\label{fig:cls}
\end{figure}

SZ clusters are typically unresolved in the \wmap\ beam, so we
treat them as point sources on the sky.  In this limit, if cluster
$i$ has flux $S_i$ at sky position $\hat{r}_i$, the sky
temperature $\Delta T(\hat{r})$ may be written as \be \Delta
T(\hat{r}) = c \sum_i S_i \delta(\hat{r},\hat{r}_i) + \sum_{\ell
m} a_{\ell m} Y_{\ell m}(\hat{r}), \label{deltaT22} \ee where
$\delta$ is the Dirac delta function, and $c$ is the conversion
factor between flux and temperature, given by \be c = c_{\ast}
\frac{\left[2\sinh(\frac{\eta}{2})\right]^2}{\eta^4}, \text{ }
c_{\ast} = \frac{1}{2k_{\rm{B}}}\left(\frac{hc}{kT_{\rm
CMB}}\right)^2 \simeq \frac{10\,\textrm{mK}}{{\rm MJy sr^{-1}}},
\label{cequ} \ee where $\eta$ is defined in Eq.~(\ref{xdefine})
(see also \citealt{Tegmark96,Tegmark98}).  Here, the spherical
harmonic coefficients $a_{\ell m}$ characterize the true CMB
temperature fluctuation.  The sky signal convolved with the beam
response, $B_{\ell}$, is \ba
\Delta T^{\rm obs}(\hat{r}) & = & c \sum_i S_i \left( \sum_{\ell} \frac{2 \ell+1}{4\pi} \, P_{\ell}(\hat{r}\cdot\hat{r}_i) \, B_{\ell} \right) \nonumber \\
 & + & \sum_{\ell m} a^{\rm tot}_{\ell m} Y_{\ell m}(\hat{r}).
\label{deltaT2} \ea Here $a^{\rm tot}_{\ell m}$ encodes the
temperature due to CMB fluctuations (convolved with the beam), and
detector noise (not convolved with the beam), \be a^{\rm
tot}_{\ell m} = a^{\rm CMB}_{\ell m} B_{\ell} + n_{\ell m},
\label{almtot} \ee where the beam transfer function, $B_{\ell}$,
is obtained from the \cite{Lambdaweb}. [Note: in this analysis, we
use the \textit{WMAP} 7-yr W-band map supplied by the \wmap\ team
(Fig.~\ref{fig:maps} upper-left), and impose the \wmap\ 7-yr
analysis mask (Fig.~\ref{fig:maps} upper-right); we also neglect
residual foreground signals in the cleaned, masked maps.]
Fig.~\ref{fig:cls} compares the angular power spectrum,
$C_{\ell}$, directly estimated from the map with the spectrum
predicted by Eq.~(\ref{almtot}) using the best-fit \lcdm\ model.
The predicted spectrum agrees with the measured spectrum very
well.

To maximize our sensitivity to point sources, we further convolve
the observed map $\Delta T^{\rm obs}$ with an optimal filter
$W_{\ell}$, so that \ba
\Delta \tilde{T}(\hat{r}) & = & c \sum_i S_i \left(\sum_{\ell} \frac{2 \ell +1}{4\pi} \, P_{\ell}(\hat{r} \cdot \hat{r}_i) \, B_{\ell} \, W_{\ell} \right) \nonumber \\
& + & \sum_{\ell m} a^{\rm tot}_{\ell m} \, W_{\ell} \, Y_{\ell
m}(\hat{r}). \label{deltaT3} \ea Note that this form implicitly
assumes that the beam response and optimal filter are both
azimuthally symmetric.  Treating the second line in
Eq.~(\ref{deltaT3}) as the noise term, we seek the form of
$W_{\ell}$ that maximizes the cluster signal-to-noise ratio.  In
the direction of cluster $i$, the filtered signal is $\Delta
\tilde{T}(\hat{r}_i) = A S_{i}$, where \be A \equiv c \sum_{\ell}
\frac{2 \ell +1}{4\pi} \, B_{\ell} \, W_{\ell} =
\textrm{constant.} \label{Aconst} \ee We choose $W_{\ell}$ to
minimize the ratio
\begin{equation}
\sigma^{2} = \textrm{Var} \left(\frac{\Delta
\tilde{\textit{T}}}{\textit{A}} \right) = \frac{\sum_{\ell}
\frac{2 \ell +1}{4\pi} \, C^{\rm tot}_{\ell} \, W^2_{\ell}}{c^2
\left(\sum_{\ell} \frac{2 \ell +1}{4\pi} \, B_{\ell} \, W_{\ell}
\right)^2}, \label{sigma2}
\end{equation}
where $C^{\rm
tot}_{\ell} \equiv B^2_{\ell} C^{\rm CMB}_{\ell} + N_{\ell}$, and
we take $C^{\rm CMB}_{\ell}$ to be the \lcdm\ model power
spectrum.

\begin{figure}
\centerline{\includegraphics[bb=0 0 650
416,width=3.2in]{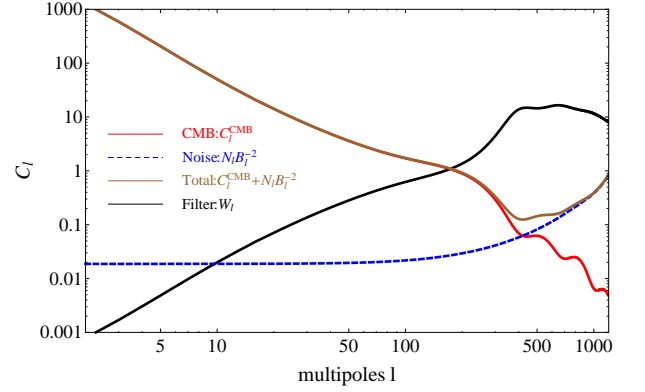}} \caption{Optimal filter (black)
for point source detection in the \wmap\ 7-yr W-band map,
Eq.~(\ref{wlfunc}).  The CMB signal (red line) and detector noise
(dashed blue line) are shown for comparison, along with their sum
(brown line).} \label{fig:filterfunc}
\end{figure}

We minimize Eq.~(\ref{sigma2}) by adding a Lagrange multiplier to
the numerator, \be \sum_{\ell} \frac{2 \ell +1}{4\pi} \, C^{\rm
tot}_{\ell} \, W^2_{\ell} + \lambda \left(\sum_{\ell} \frac{2 \ell
+1}{4\pi} \, B_{\ell} \, W_{\ell} \right)^2, \ee and minimizing
with respect $W_{\ell}$. We obtain \be W_{\ell} \sim
\frac{B_{\ell}}{B^2_{\ell} C^{\rm CMB}_{\ell} + N_{\ell}} =
\frac{B_{\ell}}{C^{\rm tot}_{\ell}}. \label{wlfunc} \ee The
normalization of $W_{\ell}$ does not affect the signal-to-noise
ratio of the cluster detection. We plot $W_{\ell}$ in
Fig.~\ref{fig:filterfunc}; note that $W_{\ell}$ is maximal in the
range $\ell \sim 500-1000$.  With this filter, the smallest
variance we can obtain for a point source is \be \sigma^2 = c^{-2}
\left( \sum_{\ell} \frac{2 \ell +1}{4\pi} \,
\frac{B^2_{\ell}}{C^{\rm tot}_{\ell}} \right)^{-1}. \ee

\subsection{The filtered cluster profile}
\label{filterpro}

What is the shape of the universal cluster profile after
filtering? Let the unfiltered temperature map due to $N$ clusters
at positions $\hat{r}_i$ ($i=1,\ldots,N$) be \be x(\hat{r}) =
\sum_i f_i(\Theta_i), \label{map1} \ee where $f_i$ is the profile
of the $i$th cluster, and $\Theta_i$ is the angle between the
$i$th cluster and $\hat{r}$, \be \cos \Theta_i = \hat{r}_i \cdot
\hat{r} = \cos\theta_i \cos\theta + \sin\theta_i \sin\theta
\cos(\phi - \phi_i). \label{costheta} \ee
The filtered cluster map may be written as
\ba
\tilde{x}(\hat{r}) & = & (W*x)(\hat{r}) \nonumber \\
%
& = & \sum_i \left[ \int d\Omega' \, f_i(\Theta') \,
W(\cos\Theta') \right], \label{filtermap1} \ea where \be
W(\cos\Theta') = \sum_{\ell} \frac{2 \ell +1}{4\pi} \, W_{\ell} \,
P_{\ell}(\cos\Theta'), \ee and $\cos\Theta' = \hat{r} \cdot
\hat{r}'$. In the limit that the SZ clusters can be considered
point sources, $f_i(\Theta_i) = c S_i \delta(\hat{r}_i,\hat{r})$,
the filtered map reduces to \be \tilde{x}(\hat{r}) = c \sum_i S_i
\sum_{\ell} \frac{2 \ell +1}{4\pi} \, W_{\ell} \,
P_{\ell}(\cos\Theta_i). \ee

\begin{figure*}
\centerline{
\includegraphics[bb=0 0 720 460,width=3.2in]{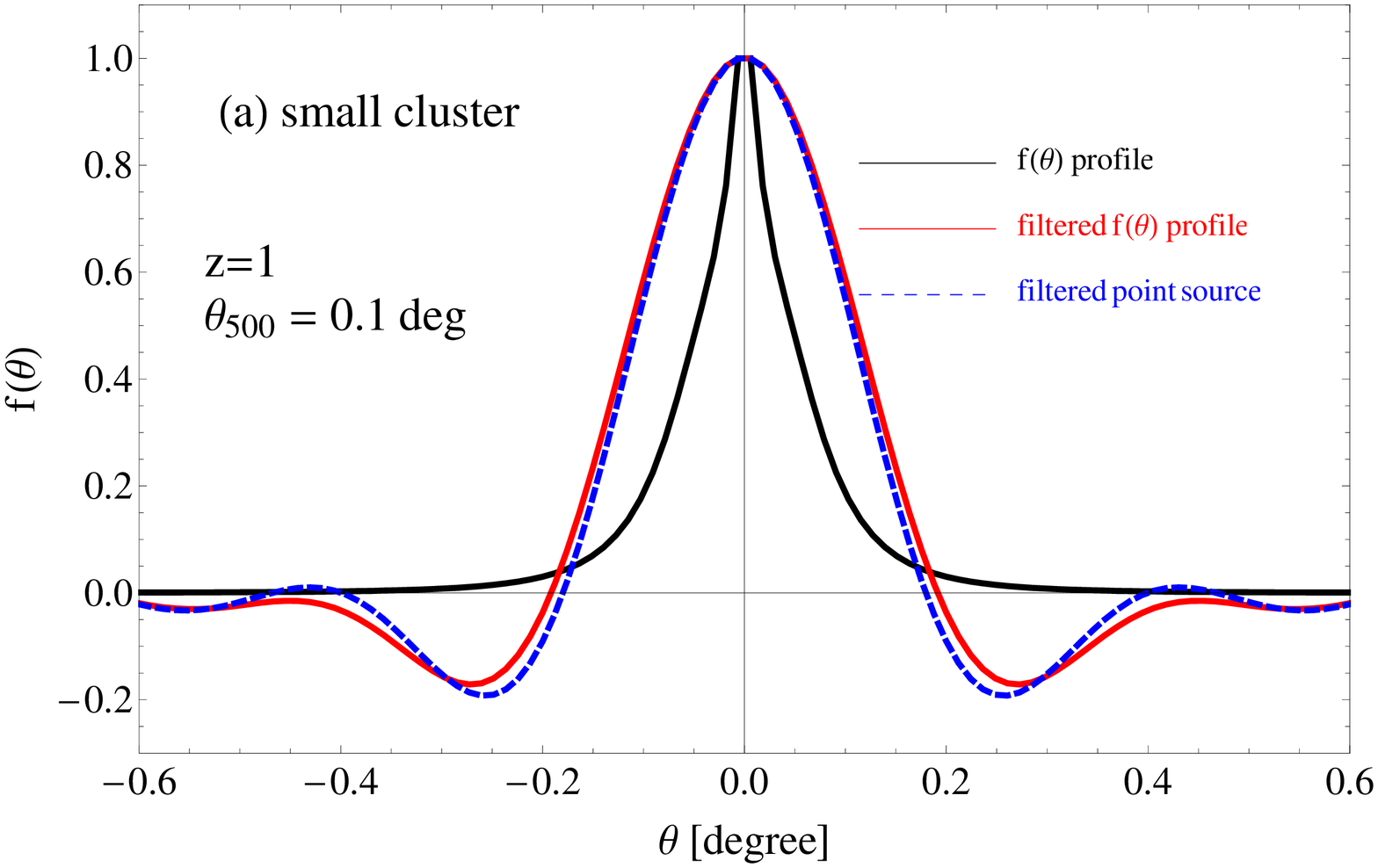}
\includegraphics[bb=0 0 687 443,width=3.2in]{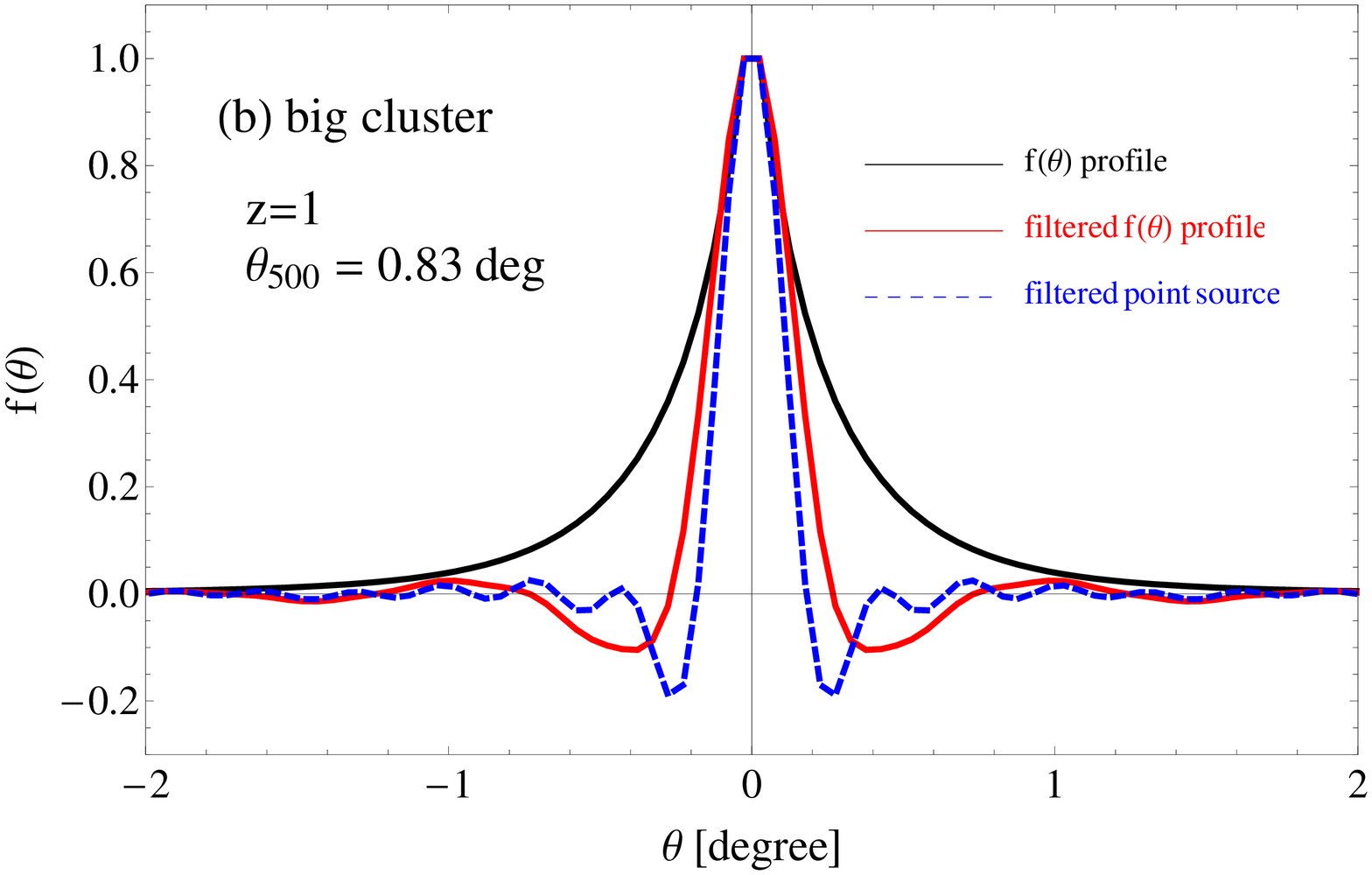}}
\caption{Shape of selected cluster profiles before and after
filtering. \textit{Left} -- a small cluster with $\theta_{500}$
less than the \wmap\ W-band beam size, $0.22^{\circ}$;
\textit{Right} -- a marginally resolved cluster.}
\label{fig:filterpro}
\end{figure*}

In Fig.~\ref{fig:filterpro}, we plot selected SZ cluster profiles
before and after filtering.  For the small cluster case shown in
Fig.~\ref{fig:filterpro}a, the filtered profile does not differ
appreciably from a filtered point source, because the cluster is
unresolved. However, for larger cluster
(Fig.~\ref{fig:filterpro}b), the filtered profile is noticeably
different from a filtered point source profile.  In either case,
we note that $\theta_{500}$ lies within the radius where the
filtered profile is still positive, so this filter should not
suppress actual SZ signal.  We consider this question in more
detail in the following section.

\subsection{Results of filtering}
\label{result}

In Fig.~\ref{fig:maps} lower-left panel, we show the filtered
\wmap\ 7-yr W-band map, which exhibits suppression of CMB signal
over a range of angular scales. The dimensionless $y$-map obtained
from $y=-\Delta T_{\rm{SZ}}/(1.56T)$ is shown in the lower-right
panel of Fig.~\ref{fig:maps}. We then calculate $Y_{500}$ for each
cluster. To do this, we use $R_{500}$ as the radius of each
cluster and sum over all of the pixels within this radius, i.e.,
\begin{eqnarray}
Y_{500}= \sum_{i}^{R \leqslant R_{500}} y(\hat{r}_{i}).
\label{Y500}
\end{eqnarray}
We then tabulate the results in column 8 of Table~\ref{tab1}. To
evaluate the uncertainty in $Y_{500}$, we simulate $1000$ sky maps
with CMB signal and pixel noise.  For each map we calculate
$Y_{500}$ at each cluster position, then compute the standard
deviation over the ensemble of maps.  The resulting error is given
in column $9$ of Table~\ref{tab1}.

\begin{figure*}
\centerline{
\includegraphics[bb=14 14 355 355,width=3.0in]{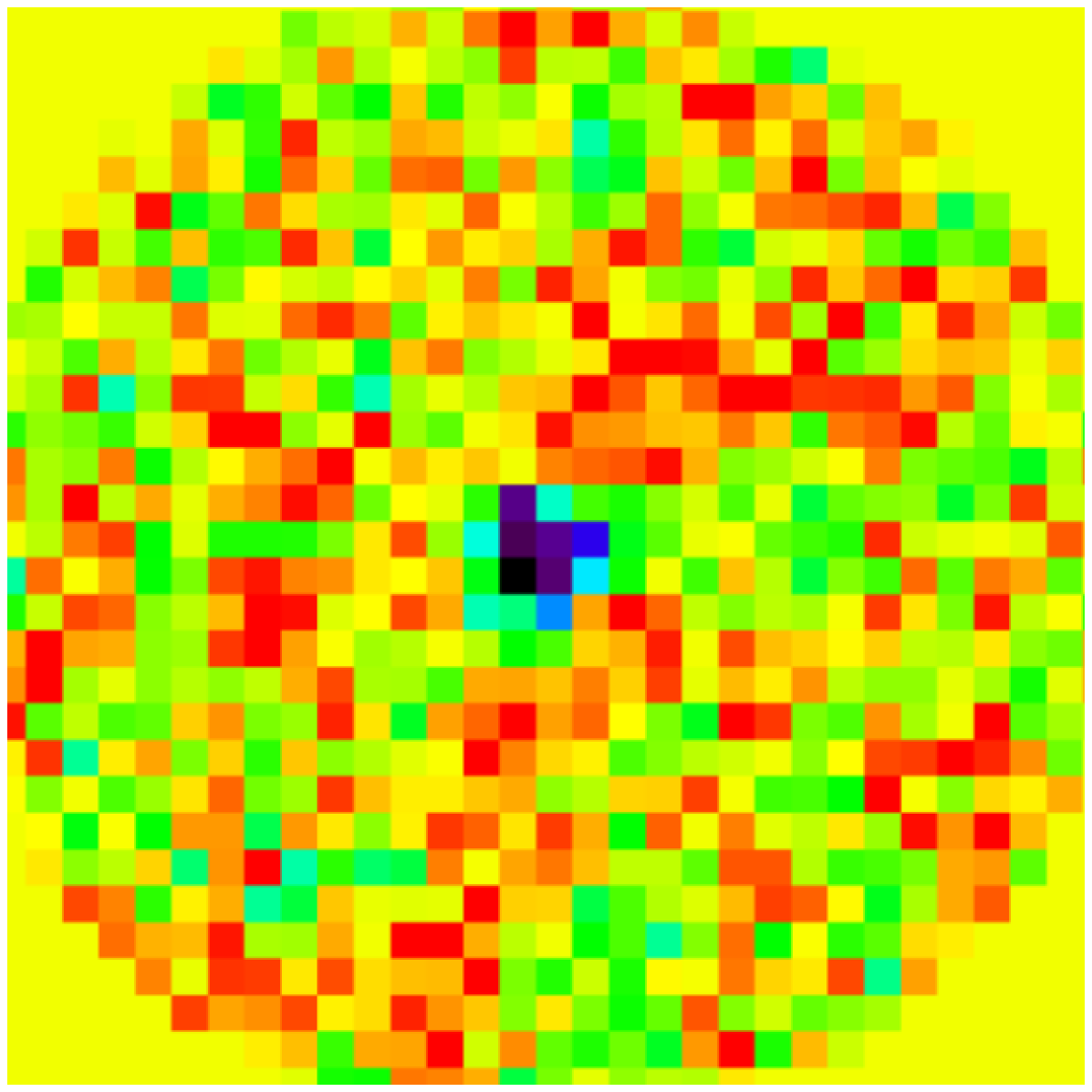}
\includegraphics[bb=14 14 355 355,width=3.0in]{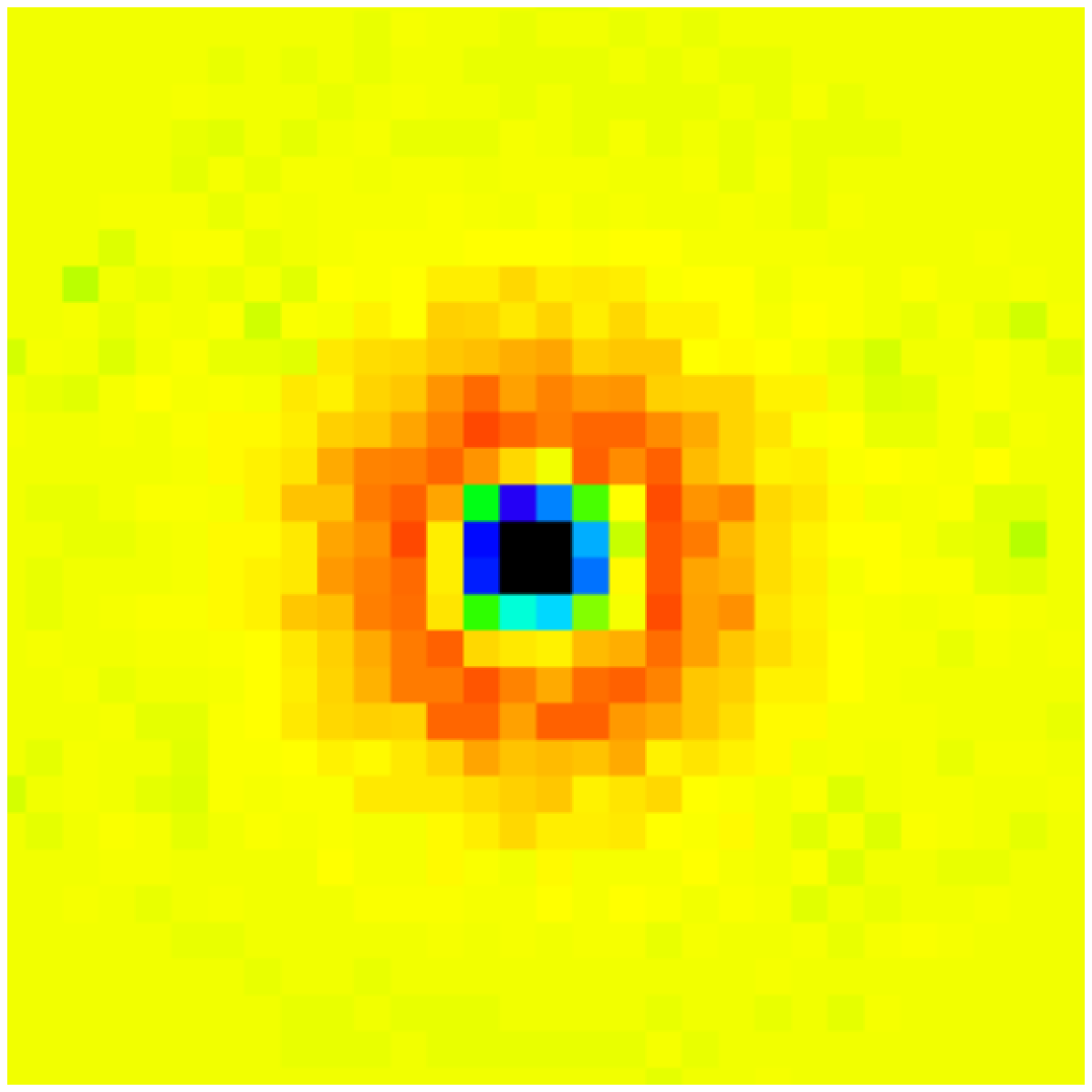}}
\caption{{\it Left} -- Average SZ profile map obtained from
stacking the \wmap\ 7-yr filtered W-band data at the locations of
the 175 SZ clusters in the \planck\ ESZ catalog. The color scale
ranges from $-4$ mK (black) to $+1$ mK (red).  {\it Right} -- The
corresponding prediction based on stacking the universal pressure
profile scaled by the \planck-measured Comptonization parameter,
$Y_{500}$. The ``Mexican-hat'' like ring is the results of
applying the matched filter.} \label{fig:sz_stack}
\end{figure*}

Because of beam dilution and detector noise in the \wmap\ data,
the signal-to-noise ratio for the detection of individual SZ
clusters is relatively low; however, the stacked signal is
detected at high significant level.  In Table~\ref{tab:stack} we
give the aggregate signal-to-noise ratio for two methods of
stacking: unweighted and weighted.   In the first method we
evaluate \be \frac{\sum_i Y_i}{\left(\sum_i \delta
Y^2_i\right)^{1/2}}, \label{unweighted} \ee which equally weights
all cluster detections.  The total signal-to-noise ratio for this
method is 8.9 for \wmap\ and 85.2 for \planck.  In the second
method we evaluate \be \left[ \sum_i \left(\frac{Y_i}{\delta
Y_i}\right)^2 \right]^{1/2}, \label{weighted} \ee which
down-weights clusters with low signal-to-noise.  This gives an
aggregate signal-to-noise of 16.3 for \wmap\ and 123.3 for
\planck.  In Fig.~\ref{fig:sz_stack} we show the aggregate SZ
profile from the stacked \wmap\ data and compare this to the
predicted profile based on stacking the universal pressure
profile, scaled by the measured Comptonization parameters from the
\planck\ catalog.

\begin{table}
\begin{centering}
\caption{Aggregate signal-to-noise for stacked SZ
clusters}\label{tab:stack}
\begin{tabular}{@{}lcc}
\hline
Method & \wmap\ & \planck\ \\
\hline
unweighted, Eq.~(\ref{unweighted}) & $8.9$ &  $85.2$ \\
weighted, Eq.~(\ref{weighted}) & $16.3$  & $123.3$ \\
\hline
\end{tabular}
\end{centering}
\end{table}


\begin{figure*}
\centerline{
\includegraphics[bb=0 0 517 332,width=3.2in]{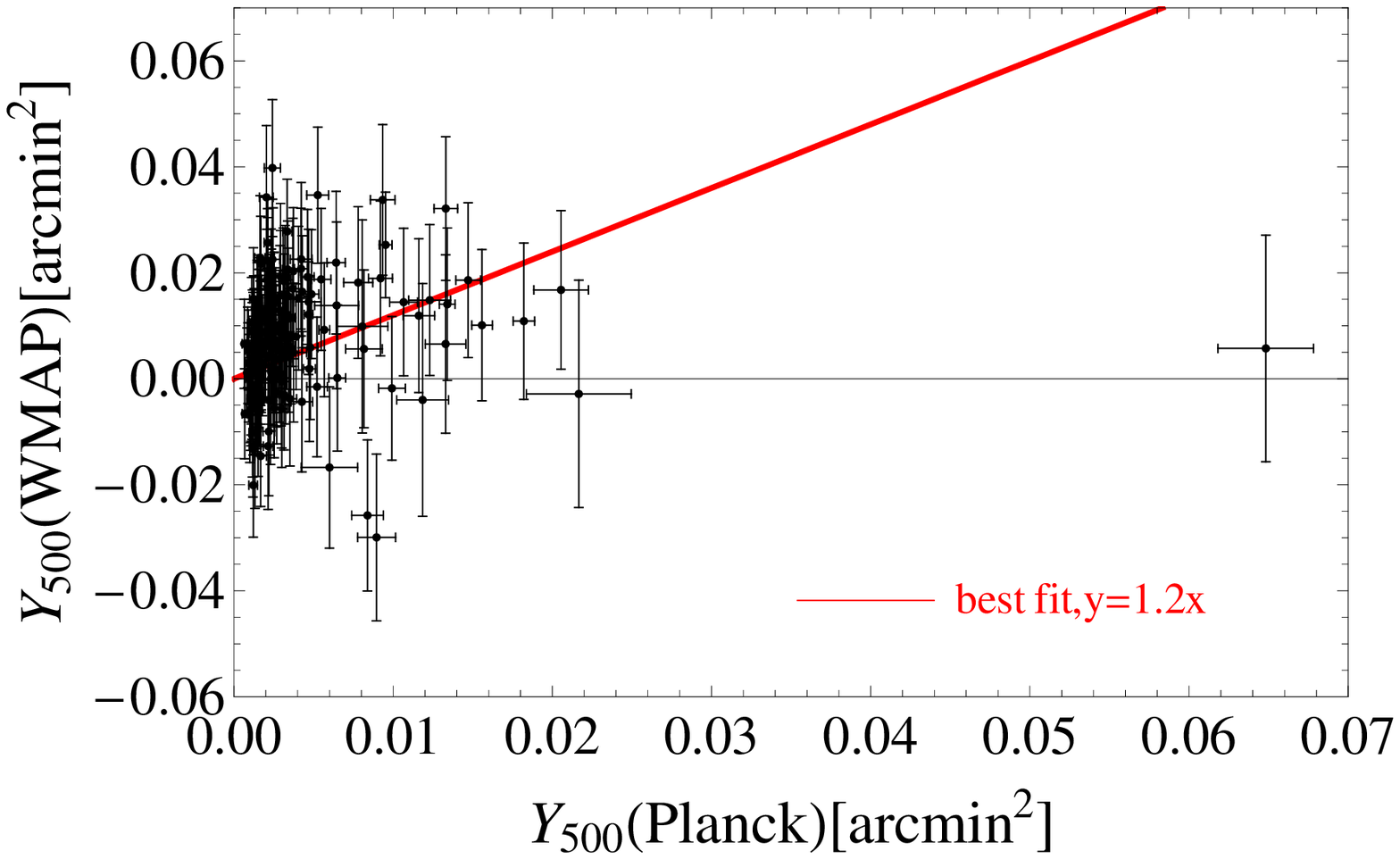}
\includegraphics[bb=0 0 517 332,width=3.2in]{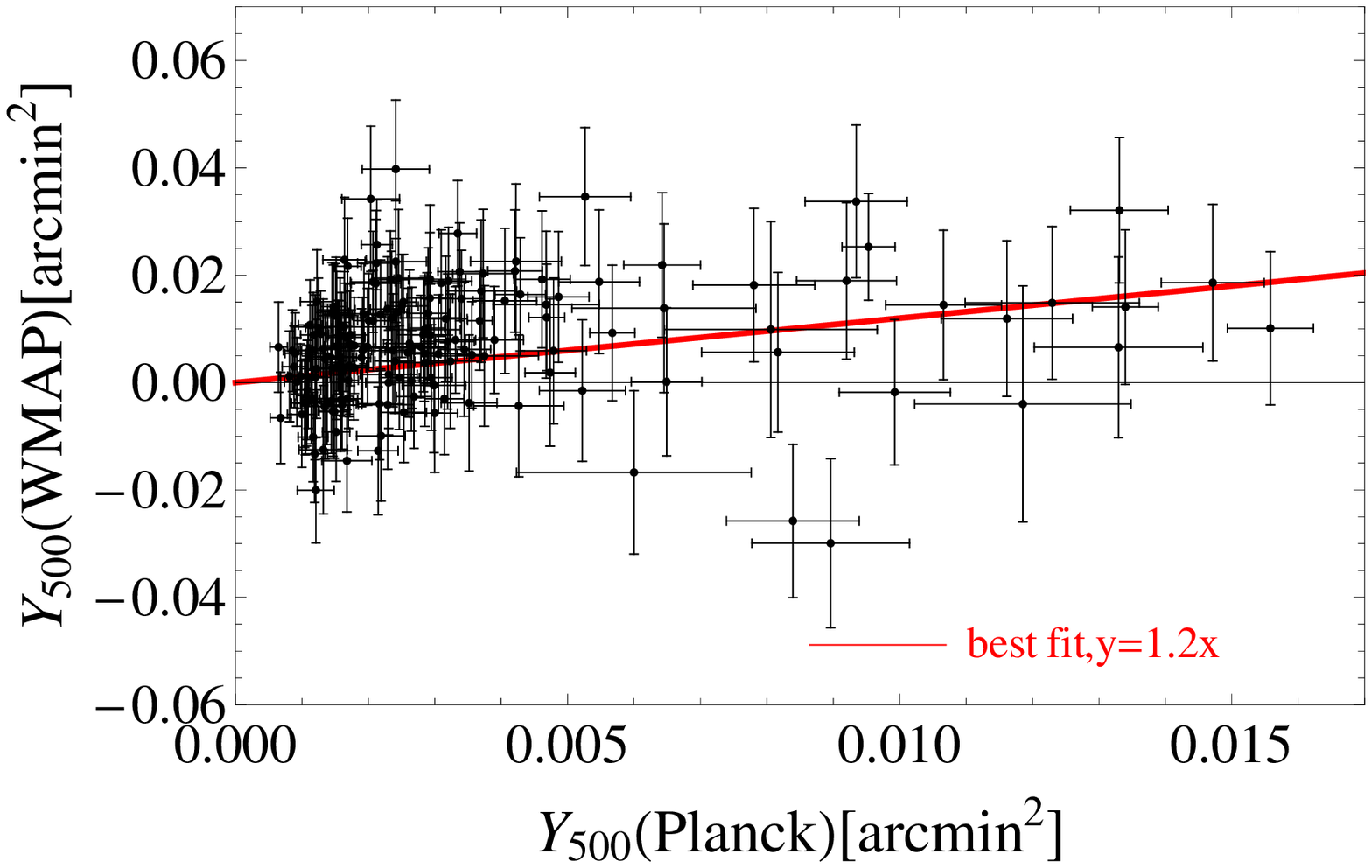}}
\caption{Comparison of 175 individual SZ cluster measurements,
$Y_{500}$, from \wmap\ and \planck.  The left panel shows the
comparison at full scale, while the right panel zooms in to a
smaller $Y_{500}$ range.  We use the relation $Y_{500} \simeq
Y_{5R500}/1.8$ to obtain $Y_{500}$ from the \planck\ catalog.}
\label{fig:yycomp}
\end{figure*}

Next, we assess the consistency between the \wmap\ and \planck\
cluster detections.  Figure~\ref{fig:yycomp} compares the
integrated Comptonization parameter, $Y_{500}$, from each data
set.  While there is quite a bit of scatter, the two measurements
are clearly correlated.  We use linear regression with errors in
both axes to fit the slope between the two data sets (see Appendix
\ref{likelidiscuss}).  Using Eq.~(\ref{pfx2}), we find \be Y^{\rm
wmap}_{500} = (1.23 \pm 0.18) \times Y^{\rm planck}_{500} \,\,
\mbox{(68\% CL),} \ee which is within $2 \sigma$ of unity.
Fig.~\ref{fig:yycomp} shows the best-fit regression in red.

While the measured slope is perfectly consistent with unity, we
briefly consider some potential sources of systematic error in our
comparison.  (1) The amplitude of $Y^{\rm wmap}_{500}$ depends on
how one treats the monopole moment in the \wmap\ map.  In our
analysis, we subtract the monopole both before and after we apply
the optimal filter to the map. (2) Finite pixel size can introduce
noise when identifying pixels within an area that is not much
larger than the pixel size. We account for this effect in our
Monte Carlo evaluation of the $Y^{\rm wmap}_{500}$ uncertainty.
(3) For clusters with $\theta_{500}$ larger than the \wmap\ W-band
beam width of $0.21^{\circ}$, the integral for $Y_{500}$ includes
some negative regions in the filtered profile, as shown in
Fig.~\ref{fig:filterpro}.  This will suppress the integrated
signal in those clusters.  The smaller value of $Y^{\rm
wmap}_{500}$ measured in the Coma cluster, compared to $Y^{\rm
planck}_{500}$, is due to this effect.  We note that our value is
consistent with that reported by
\cite{Komatsu11}\footnote{\cite{Komatsu11} do not give an explicit
value for $Y_{500}$, but from the measured dimensionless quantity
$y \sim 7\times 10^{-5}$ (two equations after their
equation~(71)), and the profile width of $10.3'$ (three lines
above their equation~(70) and figure 14), we estimate that their
measured value of $Y_{500}$ for COMA would be about $0.01$
arcmin$^2$, which is consistent with our finding, within the
errors.}.  (4) While the \planck\ team documented their detection
of SZ clusters \citep{Planck11a,Planck11b,Planck11c}, some
processing details are not explicitly described, such as the
precise choice of filter. Despite these uncertainties, the
best-fit slope between $Y^{\rm wmap}_{500}$ and $Y^{\rm
planck}_{500}$ is consistent with unity and suggests that such
effects are not significant.

\section{Conclusion}
\label{conclude}

The \planck\ collaboration released its ESZ cluster catalog in Jan
2011.  It contained $189$ SZ clusters across the full sky, 175 of
which had tabulated redshifts. In this paper, we examine these 175
cluster locations in the \wmap\ data to assess consistency with
the \planck\ catalog.

We assume that the clusters are described by a universal pressure
profile, Eq.~(\ref{unipres}), and project the 3-D profile onto the
plane of the sky.  Given this profile, we can calculate the
integrated Comptonization parameter, $Y_{500}$, for any cluster
location determined by \planck.  We filter the \wmap\ 7-yr W-band
data with an optimal filter to suppress primary CMB signal and
detector noise, then examine the effects of the filter on the
universal pressure profile.  We conclude that the angular radius,
$\theta_{500}$, adequately captures the bulk of the SZ signal in
the filtered data.   We estimate the uncertainty in $Y_{500}$
using 1000 Monte Carlo realizations of a \lcdm\ CMB signal plus
detector noise.

We perform two consistency tests between $Y^{\rm wmap}_{500}$ and
$Y^{\rm planck}_{500}$.  First, we stack all the cluster data to
estimate the total signal-to-noise ratio.  In the weighted
stacking, we obtain an aggregate signal-to-noise ratio of 16.3
from the \wmap\ data, which clearly indicates that \wmap\ detects
the most significant clusters seen by \planck.

Next we compare $Y^{\rm wmap}_{500}$ and $Y^{\rm planck}_{500}$
for each cluster. In a linear regression analysis, which accounts
for errors in both measurements, we find the best-fit slope is
$1.23 \pm 0.18$. This is consistent with unity at the $2\sigma$
confidence level.  We further consider some systematic error
sources that could lead to a slope slightly greater than unity.
Our results show that there are no fundamental problems with
reliability of the ESZ or the calibration of the SZ amplitude. A
similar conclusion was also drawn in the third version of
\citet{Witbourn11}.

The filtering technique presented here could easily be extended to other surveys, such as ACT and SPT.  We plan to revisit this issue when the appropriate data are publicly available.

\section*{Acknowledgments}
The authors acknowledge the use of the LAMBDA Data Center
(http://lambda.gsfc.nasa.gov/).  We thank Kris Sigurdson and Jim
Zibin for helpful discussions. This research was supported by the
Natural Sciences and Engineering Research Council of Canada. Y.Z.M
is supported by a CITA National Fellowship. G.H. is supported, in
part, by the Canadian Institute for Advanced Research.

\begin{appendix}

\section{Likelihood function for data with both x and y errors}
\label{likelidiscuss}

\begin{figure}
\centerline{
\includegraphics[bb=0 0 597 391,width=4.2in, keepaspectratio=false]{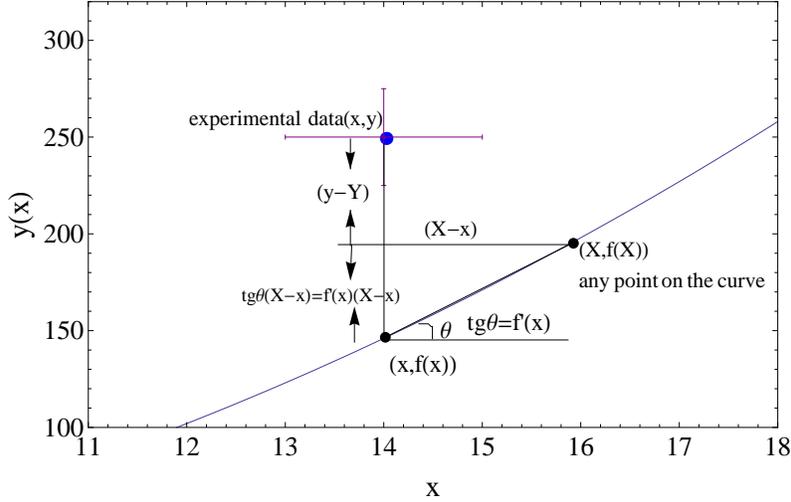}}
\caption{Geometric relation between any point on a given function
($X,f(X)$), and experimental data point ($x,y$) with error
($\delta x, \delta y$).} \label{likeliplot}
\end{figure}

In the comparison of $Y^{\rm{wmap}}$ and $Y^{\rm{planck}}$, we
would like to apply a linear regression model for these two
quantities. Since both directions have their errors, we need to
consider two-dimensional errors when fitting of the slope
parameter (see also \citealt{Clutton-Brock67}, Section 15.3 of
\citealt{Press07} and \citealt{Akritas96}).

Suppose we have one data point $(x,y)$, as plotted in
Fig.~\ref{likeliplot}, with measurement errors in both directions
$(\delta x,\delta y)$. The function we want to fit is $Y=f(X)$,
and so $(X,Y)$ can be any point on the curve. In practise, $(X,Y)$
should be fairly close to the data point ($x,y$). We want to
measure \ba
y-f(x) &=&(y-Y)+(Y-f(x))  \nonumber \\
&=&(y-Y)+(f(X)-f(x))  \nonumber \\
&=&(y-Y)+\tan \theta (X-x)  \nonumber \\
&=&(y-Y)+f^{\prime }(x)(X-x).
\ea%
The geometric relationship between these quantities is illustrated in Fig.~\ref%
{likeliplot}. Therefore,
\ba
\textrm{Var}(y-f(x)) &=& \textrm{Var}(y-Y)+f^{\prime }(x)^{2} \textrm{Var}((X-x))  \nonumber \\
&=&\delta y^{2}+f^{\prime }(x)^{2}\delta x^{2}.
\ea%
Therefore, for just one datum, the likelihood for the function is
$f(x)$ as \be \mathcal{L}(f(x))=\frac{1}{\left[ 2\pi \left( \delta
y^{2}+f^{\prime }(x)^{2}\delta
x^{2}\right) \right] ^{\frac{1}{2}}}\exp \left[ -\frac{1}{2}\left( \frac{%
(y-f(x))^{2}}{\left( \delta y^{2}+f^{\prime }(x)^{2}\delta x^{2}\right) }%
\right) \right] .
\ee%
We can generalize above equation for a data set with $N$
independent data ($x_{k},y_{k},\delta x_{k},\delta y_{k}$),
yielding \be
\mathcal{L}(f(x))=\prod\limits_{k=1}^{N}\frac{1}{\left[ 2\pi
\left( \delta
y_{k}^{2}+f^{\prime }(x_{k})^{2}\delta x_{k}^{2}\right) \right] ^{\frac{1}{2}%
}}\exp \left[ -\frac{1}{2}\left(
\frac{(y_{k}-f(x_{k}))^{2}}{\left( \delta y_{k}^{2}+f^{\prime
}(x_{k})^{2}\delta x_{k}^{2}\right) }\right) \right] .
\label{pfx1}
\ee

Therefore, for the case of a linear regression model $y=a*x$, the
likelihood for parameter $a$ becomes
\be
\mathcal{L}(a)=\prod\limits_{k=1}^{N}\frac{1}{\left[ 2\pi \left(
\delta
y_{k}^{2}+(a \delta x_{k})^{2}\right) \right] ^{\frac{1}{2}%
}}\exp \left[ -\frac{1}{2}\left( \frac{(y_{k}-a*x_{k})^{2}}{\left(
\delta y_{k}^{2}+(a\delta x_{k})^{2}\right) }\right) \right] .
\label{pfx2}
\ee

We maximize this to obtain the amplitude of the linear regression
parameter between $Y^{\rm{wmap}}$ and $Y^{\rm{planck}}$ in
Section~\ref{filter}.

\begin{table}
\begin{centering}
\caption{\wmap\ observations of \planck\ ESZ clusters}
\label{tab1}
\begin{tabular}{@{}lrrrrrrrr}
\hline
Index & Gal lon ($l$) & Gal lat ($b$) & Redshift & $\theta_{500}$ & $Y^{\rm planck}_{500}$ & $\delta Y^{\rm planck}_{500}$ & $Y^{\rm wmap}_{500}$ & $\delta Y^{\rm wmap}_{500}$ \\
 & (deg) & (deg) & & (arcmin) & (arcmin$^2$) & (arcmin$^2$) & (arcmin$^2$) & (arcmin$^2$) \\
\hline
   1 &   $0.4409$ &$-41.8351$ &   $0.1651$ &  $6.3071$ &   $0.14 \times 10^{-2}$ &   $0.27 \times 10^{-3}$ &   $0.48 \times 10^{-2}$ &   $0.92 \times 10^{-2} $\\
   2 &$   2.7497 $&$ -56.1827 $&$   0.1411 $&$   7.1114 $&$   0.17 \times 10^{-2} $&$   0.27 \times 10^{-3} $&$   0.71 \times 10^{-2} $&$   0.10 \times 10^{-1}$ \\
   3 &$   3.9081 $&$ -59.4171 $&$   0.1510 $&$   7.6784 $&$   0.33 \times 10^{-2} $&$   0.28 \times 10^{-3} $&$   0.27 \times 10^{-1} $&$   0.10 \times 10^{-1}$ \\
   4 &$   6.4760 $&$  50.5490 $&$   0.0766 $&$  15.5982 $&$   0.99 \times 10^{-2} $&$   0.84 \times 10^{-3} $&$  -0.18 \times 10^{-2} $&$   0.14 \times 10^{-1}$ \\
   5 &$   6.7046 $&$ -35.5407 $&$   0.0894 $&$  10.5796 $&$   0.32 \times 10^{-2} $&$   0.44 \times 10^{-3} $&$   0.39 \times 10^{-2} $&$   0.13 \times 10^{-1}$ \\
   6 &$   6.7849 $&$  30.4683 $&$   0.2030 $&$   7.5616 $&$   0.95 \times 10^{-2} $&$   0.40 \times 10^{-3} $&$   0.25 \times 10^{-1} $&$   0.10 \times 10^{-1}$ \\
   7 &$   8.3005 $&$ -64.7564 $&$   0.3120 $&$   8.7664 $&$   0.27 \times 10^{-2} $&$   0.56 \times 10^{-3} $&$   0.63 \times 10^{-2} $&$   0.11 \times 10^{-1}$ \\
   8 &$   8.4486 $&$ -56.3564 $&$   0.1486 $&$   6.7883 $&$   0.13 \times 10^{-2} $&$   0.26 \times 10^{-3} $&$  -0.44 \times 10^{-2} $&$   0.97 \times 10^{-2}$ \\
   9 &$   8.9362 $&$ -81.2386 $&$   0.3066 $&$   4.6335 $&$   0.23 \times 10^{-2} $&$   0.27 \times 10^{-3} $&$   0.19 \times 10^{-1} $&$   0.87 \times 10^{-2}$ \\
  10 &$  18.5314 $&$ -25.7228 $&$   0.3171 $&$   4.1424 $&$   0.15 \times 10^{-2} $&$   0.22 \times 10^{-3} $&$  -0.52 \times 10^{-2} $&$   0.84 \times 10^{-2}$ \\
  11 &$  21.0918 $&$  33.2560 $&$   0.1514 $&$   9.0695 $&$   0.42 \times 10^{-2} $&$   0.41 \times 10^{-3} $&$   0.20 \times 10^{-1} $&$   0.11 \times 10^{-1}$ \\
  12 &$  29.0054 $&$  44.5625 $&$   0.0353 $&$  21.7792 $&$   0.82 \times 10^{-2} $&$   0.12 \times 10^{-2} $&$   0.56 \times 10^{-2} $&$   0.15 \times 10^{-1}$ \\
  13 &$  33.4614 $&$ -48.4318 $&$   0.0943 $&$   8.8429 $&$   0.30 \times 10^{-2} $&$   0.34 \times 10^{-3} $&$  -0.56 \times 10^{-2} $&$   0.11 \times 10^{-1}$ \\
  14 &$  33.7814 $&$  77.1628 $&$   0.0622 $&$  17.1573 $&$   0.93 \times 10^{-2} $&$   0.77 \times 10^{-3} $&$   0.33 \times 10^{-1} $&$   0.13 \times 10^{-1}$ \\
  15 &$  36.7219 $&$  14.9232 $&$   0.1525 $&$   7.0561 $&$   0.19 \times 10^{-2} $&$   0.34 \times 10^{-3} $&$   0.58 \times 10^{-2} $&$   0.98 \times 10^{-2}$ \\
  16 &$  39.8595 $&$ -39.9889 $&$   0.1760 $&$   5.9780 $&$   0.17 \times 10^{-2} $&$   0.29 \times 10^{-3} $&$   0.30 \times 10^{-2} $&$   0.94 \times 10^{-2}$ \\
  17 &$  42.8257 $&$  56.6172 $&$   0.0723 $&$  12.6087 $&$   0.55 \times 10^{-2} $&$   0.61 \times 10^{-3} $&$   0.18 \times 10^{-1} $&$   0.13 \times 10^{-1}$ \\
  18 &$  44.2253 $&$  48.6842 $&$   0.0894 $&$  14.0653 $&$   0.13 \times 10^{-1} $&$   0.74 \times 10^{-3} $&$   0.32 \times 10^{-1} $&$   0.13 \times 10^{-1}$ \\
  19 &$  46.0815 $&$  27.1816 $&$   0.3890 $&$   3.5573 $&$   0.12 \times 10^{-2} $&$   0.19 \times 10^{-3} $&$   0.87 \times 10^{-2} $&$   0.76 \times 10^{-2}$ \\
  20 &$  46.5037 $&$ -49.4389 $&$   0.0846 $&$  10.8945 $&$   0.35 \times 10^{-2} $&$   0.42 \times 10^{-3} $&$  -0.38 \times 10^{-2} $&$   0.13 \times 10^{-1}$ \\
  21 &$  46.8830 $&$  56.4988 $&$   0.1145 $&$   9.0824 $&$   0.37 \times 10^{-2} $&$   0.46 \times 10^{-3} $&$   0.20 \times 10^{-1} $&$   0.11 \times 10^{-1}$ \\
  22 &$  48.0522 $&$  57.1769 $&$   0.0777 $&$  10.9294 $&$   0.37 \times 10^{-2} $&$   0.54 \times 10^{-3} $&$   0.17 \times 10^{-1} $&$   0.13 \times 10^{-1}$ \\
  23 &$  49.2044 $&$  30.8618 $&$   0.1644 $&$   7.2635 $&$   0.24 \times 10^{-2} $&$   0.30 \times 10^{-3} $&$   0.39 \times 10^{-2} $&$   0.97 \times 10^{-2}$ \\
  24 &$  49.3365 $&$  44.3823 $&$   0.0972 $&$   8.5539 $&$   0.30 \times 10^{-2} $&$   0.48 \times 10^{-3} $&$  -0.52 \times 10^{-3} $&$   0.11 \times 10^{-1}$ \\
  25 &$  49.6670 $&$ -49.5094 $&$   0.0980 $&$   9.5536 $&$   0.21 \times 10^{-2} $&$   0.37 \times 10^{-3} $&$   0.18 \times 10^{-1} $&$   0.12 \times 10^{-1}$ \\
  26 &$  53.4434 $&$ -36.2698 $&$   0.3250 $&$   1.6148 $&$   0.98 \times 10^{-3} $&$   0.19 \times 10^{-3} $&$   0.12 \times 10^{-2} $&$   0.59 \times 10^{-2}$ \\
  27 &$  53.5227 $&$  59.5442 $&$   0.1130 $&$   8.8081 $&$   0.31 \times 10^{-2} $&$   0.43 \times 10^{-3} $&$   0.53 \times 10^{-2} $&$   0.11 \times 10^{-1}$ \\
  28 &$  55.6010 $&$  31.8644 $&$   0.2240 $&$   6.1481 $&$   0.27 \times 10^{-2} $&$   0.26 \times 10^{-3} $&$  -0.26 \times 10^{-2} $&$   0.99 \times 10^{-2}$ \\
  29 &$  55.9755 $&$ -34.8850 $&$   0.1244 $&$   6.7696 $&$   0.20 \times 10^{-2} $&$   0.28 \times 10^{-3} $&$   0.65 \times 10^{-2} $&$   0.98 \times 10^{-2}$ \\
  30 &$  56.8121 $&$  36.3165 $&$   0.0953 $&$  10.6714 $&$   0.32 \times 10^{-2} $&$   0.40 \times 10^{-3} $&$   0.75 \times 10^{-2} $&$   0.12 \times 10^{-1}$ \\
  31 &$  56.9685 $&$ -55.0798 $&$   0.4470 $&$   3.5090 $&$   0.16 \times 10^{-2} $&$   0.20 \times 10^{-3} $&$   0.86 \times 10^{-2} $&$   0.76 \times 10^{-2}$ \\
  32 &$  57.2694 $&$ -45.3577 $&$   0.3970 $&$   4.3122 $&$   0.18 \times 10^{-2} $&$   0.21 \times 10^{-3} $&$   0.68 \times 10^{-2} $&$   0.86 \times 10^{-2}$ \\
  33 &$  57.3362 $&$  88.0113 $&$   0.0231 $&$  40.6731 $&$   0.65 \times 10^{-1} $&$   0.30 \times 10^{-2} $&$   0.56 \times 10^{-2} $&$   0.21 \times 10^{-1}$ \\
  34 &$  57.6138 $&$  34.9421 $&$   0.0802 $&$  10.4783 $&$   0.29 \times 10^{-2} $&$   0.39 \times 10^{-3} $&$   0.37 \times 10^{-2} $&$   0.13 \times 10^{-1}$ \\
  35 &$  57.9289 $&$  27.6442 $&$   0.0757 $&$  11.7297 $&$   0.20 \times 10^{-2} $&$   0.43 \times 10^{-3} $&$   0.34 \times 10^{-1} $&$   0.13 \times 10^{-1}$ \\
  36 &$  58.2818 $&$  18.5938 $&$   0.0650 $&$  12.5308 $&$   0.48 \times 10^{-2} $&$   0.50 \times 10^{-3} $&$   0.58 \times 10^{-2} $&$   0.13 \times 10^{-1}$ \\
  37 &$  62.4239 $&$ -46.4150 $&$   0.0906 $&$   9.4655 $&$   0.23 \times 10^{-2} $&$   0.39 \times 10^{-3} $&$  -0.41 \times 10^{-2} $&$   0.12 \times 10^{-1}$ \\
  38 &$  62.9261 $&$  43.7096 $&$   0.0299 $&$  27.6348 $&$   0.13 \times 10^{-1} $&$   0.13 \times 10^{-2} $&$   0.65 \times 10^{-2} $&$   0.16 \times 10^{-1}$ \\
  39 &$  67.2317 $&$  67.4641 $&$   0.1712 $&$   7.4168 $&$   0.31 \times 10^{-2} $&$   0.30 \times 10^{-3} $&$  -0.30 \times 10^{-2} $&$   0.10 \times 10^{-1}$ \\
  40 &$  71.6149 $&$  29.7983 $&$   0.1565 $&$   5.4367 $&$   0.13 \times 10^{-2} $&$   0.20 \times 10^{-3} $&$   0.62 \times 10^{-2} $&$   0.92 \times 10^{-2}$ \\
  41 &$  72.6308 $&$  41.4639 $&$   0.2280 $&$   6.2793 $&$   0.47 \times 10^{-2} $&$   0.27 \times 10^{-3} $&$   0.12 \times 10^{-1} $&$   0.97 \times 10^{-2}$ \\
  42 &$  72.8026 $&$ -18.7212 $&$   0.1430 $&$   7.4432 $&$   0.28 \times 10^{-2} $&$   0.54 \times 10^{-3} $&$   0.97 \times 10^{-2} $&$   0.96 \times 10^{-2}$ \\
  43 &$  73.9652 $&$ -27.8216 $&$   0.2329 $&$   6.2334 $&$   0.31 \times 10^{-2} $&$   0.26 \times 10^{-3} $&$   0.18 \times 10^{-1} $&$   0.96 \times 10^{-2}$ \\
  44 &$  77.9099 $&$ -26.6467 $&$   0.1470 $&$   7.2926 $&$   0.22 \times 10^{-2} $&$   0.30 \times 10^{-3} $&$   0.12 \times 10^{-1} $&$   0.10 \times 10^{-1}$ \\
  45 &$  80.3824 $&$ -33.2035 $&$   0.1072 $&$   7.8795 $&$   0.22 \times 10^{-2} $&$   0.35 \times 10^{-3} $&$  -0.39 \times 10^{-2} $&$   0.97 \times 10^{-2}$ \\
  46 &$  80.9954 $&$ -50.9072 $&$   0.2998 $&$   4.5969 $&$   0.15 \times 10^{-2} $&$   0.25 \times 10^{-3} $&$   0.13 \times 10^{-1} $&$   0.89 \times 10^{-2}$ \\
  47 &$  83.2875 $&$ -31.0322 $&$   0.4120 $&$   3.9787 $&$   0.12 \times 10^{-2} $&$   0.22 \times 10^{-3} $&$  -0.10 \times 10^{-1} $&$   0.87 \times 10^{-2}$ \\
  48 &$  85.9999 $&$  26.7107 $&$   0.1790 $&$  11.3240 $&$   0.24 \times 10^{-2} $&$   0.10 \times 10^{-2} $&$   0.12 \times 10^{-1} $&$   0.12 \times 10^{-1}$ \\
  49 &$  86.4555 $&$  15.2999 $&$   0.2600 $&$   4.9172 $&$   0.17 \times 10^{-2} $&$   0.15 \times 10^{-3} $&$   0.41 \times 10^{-3} $&$   0.91 \times 10^{-2}$ \\
  50 &$  92.7308 $&$  73.4614 $&$   0.2279 $&$   5.5586 $&$   0.25 \times 10^{-2} $&$   0.25 \times 10^{-3} $&$   0.14 \times 10^{-1} $&$   0.90 \times 10^{-2}$ \\
  51 &$  93.9197 $&$  34.9077 $&$   0.0809 $&$  11.6216 $&$   0.57 \times 10^{-2} $&$   0.34 \times 10^{-3} $&$   0.91 \times 10^{-2} $&$   0.13 \times 10^{-1}$ \\
  52 &$  94.0187 $&$  27.4256 $&$   0.2990 $&$   8.0500 $&$   0.16 \times 10^{-2} $&$   0.80 \times 10^{-3} $&$   0.10 \times 10^{-1} $&$   0.10 \times 10^{-1}$ \\
  53 &$  96.8523 $&$  52.4668 $&$   0.3179 $&$   4.1455 $&$   0.81 \times 10^{-3} $&$   0.14 \times 10^{-3} $&$   0.11 \times 10^{-2} $&$   0.83 \times 10^{-2}$ \\
  54 &$  97.7396 $&$  38.1199 $&$   0.1709 $&$   6.3915 $&$   0.24 \times 10^{-2} $&$   0.17 \times 10^{-3} $&$   0.13 \times 10^{-1} $&$   0.96 \times 10^{-2}$ \\
  55 &$  98.9502 $&$  24.8610 $&$   0.0928 $&$   8.2040 $&$   0.12 \times 10^{-2} $&$   0.23 \times 10^{-3} $&$   0.95 \times 10^{-3} $&$   0.11 \times 10^{-1}$ \\
  56 &$ 106.7311 $&$ -83.2257 $&$   0.2924 $&$   4.2902 $&$   0.20 \times 10^{-2} $&$   0.24 \times 10^{-3} $&$   0.11 \times 10^{-1} $&$   0.83 \times 10^{-2}$ \\
  57 &$ 107.1124 $&$  65.3142 $&$   0.2799 $&$   4.9453 $&$   0.17 \times 10^{-2} $&$   0.20 \times 10^{-3} $&$   0.21 \times 10^{-1} $&$   0.88 \times 10^{-2}$ \\
  58 &$ 110.9809 $&$  31.7338 $&$   0.0581 $&$  16.6276 $&$   0.13 \times 10^{-1} $&$   0.50 \times 10^{-3} $&$   0.14 \times 10^{-1} $&$   0.14 \times 10^{-1}$ \\
  59 &$ 112.4561 $&$  57.0378 $&$   0.0701 $&$  11.2395 $&$   0.29 \times 10^{-2} $&$   0.35 \times 10^{-3} $&$   0.15 \times 10^{-1} $&$   0.13 \times 10^{-1}$ \\
  60 &$ 113.8229 $&$  44.3503 $&$   0.2250 $&$   4.6399 $&$   0.65 \times 10^{-3} $&$   0.13 \times 10^{-3} $&$   0.65 \times 10^{-2} $&$   0.87 \times 10^{-2}$ \\

\hline
\end{tabular}
\end{centering}
\end{table}

\addtocounter{table}{-1}
\begin{table}
\begin{centering}
\caption{\wmap\ observations of \planck\ ESZ clusters}
\label{tab2}
\begin{tabular}{@{}lrrrrrrrr}
\hline
Index & Gal lon ($l$) & Gal lat ($b$) & Redshift & $\theta_{500}$ & $Y^{\rm planck}_{500}$ & $\delta Y^{\rm planck}_{500}$ & $Y^{\rm wmap}_{500}$ & $\delta Y^{\rm wmap}_{500}$ \\
 & (deg) & (deg) & & (arcmin) & (arcmin$^2$) & (arcmin$^2$) & (arcmin$^2$) & (arcmin$^2$) \\
\hline
  61 &$ 114.3368 $&$  64.8740 $&$   0.2836 $&$   4.1782 $&$   0.11 \times 10^{-2} $&$   0.18 \times 10^{-3} $&$   0.11 \times 10^{-1} $&$   0.85 \times 10^{-2}$ \\
  62 &$ 115.1624 $&$ -72.0911 $&$   0.0555 $&$  18.8448 $&$   0.12 \times 10^{-1} $&$   0.99 \times 10^{-3} $&$   0.12 \times 10^{-1} $&$   0.15 \times 10^{-1}$ \\
  63 &$ 118.4482 $&$  39.3351 $&$   0.3967 $&$   3.4480 $&$   0.86 \times 10^{-3} $&$   0.14 \times 10^{-3} $&$   0.58 \times 10^{-2} $&$   0.74 \times 10^{-2}$ \\
  64 &$ 118.6016 $&$  28.5586 $&$   0.1780 $&$   5.9658 $&$   0.12 \times 10^{-2} $&$   0.25 \times 10^{-3} $&$   0.15 \times 10^{-1} $&$   0.96 \times 10^{-2}$ \\
  65 &$ 124.2179 $&$ -36.4859 $&$   0.1971 $&$   6.2310 $&$   0.28 \times 10^{-2} $&$   0.38 \times 10^{-3} $&$   0.66 \times 10^{-2} $&$   0.96 \times 10^{-2}$ \\
  66 &$ 125.5865 $&$ -64.1447 $&$   0.0442 $&$  17.7296 $&$   0.78 \times 10^{-2} $&$   0.92 \times 10^{-3} $&$   0.18 \times 10^{-1} $&$   0.14 \times 10^{-1}$ \\
  67 &$ 125.7057 $&$  53.8566 $&$   0.3019 $&$   4.1887 $&$   0.10 \times 10^{-2} $&$   0.15 \times 10^{-3} $&$  -0.38 \times 10^{-2} $&$   0.87 \times 10^{-2}$ \\
  68 &$ 139.1979 $&$  56.3560 $&$   0.3220 $&$   3.7032 $&$   0.68 \times 10^{-3} $&$   0.16 \times 10^{-3} $&$  -0.65 \times 10^{-2} $&$   0.79 \times 10^{-2}$ \\
  69 &$ 143.2459 $&$  65.2152 $&$   0.2110 $&$   4.8535 $&$   0.12 \times 10^{-2} $&$   0.19 \times 10^{-3} $&$   0.10 \times 10^{-1} $&$   0.90 \times 10^{-2}$ \\
  70 &$ 146.3311 $&$ -15.5913 $&$   0.0172 $&$  39.9005 $&$   0.22 \times 10^{-1} $&$   0.33 \times 10^{-2} $&$  -0.28 \times 10^{-2} $&$   0.21 \times 10^{-1}$ \\
  71 &$ 149.2421 $&$  54.1894 $&$   0.1369 $&$   7.3566 $&$   0.29 \times 10^{-2} $&$   0.26 \times 10^{-3} $&$   0.10 \times 10^{-1} $&$   0.97 \times 10^{-2}$ \\
  72 &$ 149.7332 $&$  34.6991 $&$   0.1818 $&$   6.9874 $&$   0.33 \times 10^{-2} $&$   0.30 \times 10^{-3} $&$   0.78 \times 10^{-2} $&$   0.93 \times 10^{-2}$ \\
  73 &$ 159.8592 $&$ -73.4730 $&$   0.2060 $&$   5.6338 $&$   0.29 \times 10^{-2} $&$   0.29 \times 10^{-3} $&$   0.87 \times 10^{-2} $&$   0.97 \times 10^{-2}$ \\
  74 &$ 161.4447 $&$  26.2337 $&$   0.0381 $&$  17.7570 $&$   0.42 \times 10^{-2} $&$   0.68 \times 10^{-3} $&$   0.22 \times 10^{-1} $&$   0.14 \times 10^{-1}$ \\
  75 &$ 163.7234 $&$  53.5331 $&$   0.1580 $&$   6.4145 $&$   0.17 \times 10^{-2} $&$   0.23 \times 10^{-3} $&$  -0.30 \times 10^{-2} $&$   0.98 \times 10^{-2}$ \\
  76 &$ 164.1859 $&$ -38.8932 $&$   0.0739 $&$  14.8848 $&$   0.11 \times 10^{-1} $&$   0.87 \times 10^{-3} $&$   0.14 \times 10^{-1} $&$   0.13 \times 10^{-1}$ \\
  77 &$ 165.0876 $&$  54.1196 $&$   0.1440 $&$   7.1812 $&$   0.15 \times 10^{-2} $&$   0.26 \times 10^{-3} $&$   0.12 \times 10^{-1} $&$   0.97 \times 10^{-2}$ \\
  78 &$ 166.1319 $&$  43.3926 $&$   0.2172 $&$   5.6392 $&$   0.21 \times 10^{-2} $&$   0.24 \times 10^{-3} $&$   0.11 \times 10^{-1} $&$   0.91 \times 10^{-2}$ \\
  79 &$ 167.6563 $&$  17.6484 $&$   0.1740 $&$   6.3019 $&$   0.25 \times 10^{-2} $&$   0.30 \times 10^{-3} $&$   0.89 \times 10^{-3} $&$   0.92 \times 10^{-2}$ \\
  80 &$ 171.9474 $&$ -40.6552 $&$   0.2700 $&$   5.9117 $&$   0.34 \times 10^{-2} $&$   0.32 \times 10^{-3} $&$   0.15 \times 10^{-1} $&$   0.94 \times 10^{-2}$ \\
  81 &$ 172.8864 $&$  65.3231 $&$   0.0794 $&$   9.4305 $&$   0.16 \times 10^{-2} $&$   0.33 \times 10^{-3} $&$   0.23 \times 10^{-1} $&$   0.11 \times 10^{-1}$ \\
  82 &$ 176.2823 $&$ -35.0506 $&$   0.0347 $&$  25.2705 $&$   0.65 \times 10^{-2} $&$   0.14 \times 10^{-2} $&$   0.14 \times 10^{-1} $&$   0.16 \times 10^{-1}$ \\
  83 &$ 180.2409 $&$  21.0459 $&$   0.5460 $&$   3.5482 $&$   0.15 \times 10^{-2} $&$   0.21 \times 10^{-3} $&$   0.13 \times 10^{-1} $&$   0.77 \times 10^{-2}$ \\
  84 &$ 180.6237 $&$  76.6529 $&$   0.2138 $&$   5.0805 $&$   0.15 \times 10^{-2} $&$   0.21 \times 10^{-3} $&$  -0.91 \times 10^{-2} $&$   0.94 \times 10^{-2}$ \\
  85 &$ 182.4440 $&$ -28.2986 $&$   0.0882 $&$  13.0626 $&$   0.92 \times 10^{-2} $&$   0.75 \times 10^{-3} $&$   0.19 \times 10^{-1} $&$   0.13 \times 10^{-1}$ \\
  86 &$ 182.6361 $&$  55.8245 $&$   0.2060 $&$   5.4790 $&$   0.11 \times 10^{-2} $&$   0.23 \times 10^{-3} $&$  -0.38 \times 10^{-2} $&$   0.91 \times 10^{-2}$ \\
  87 &$ 186.3949 $&$  37.2555 $&$   0.2820 $&$   5.0003 $&$   0.28 \times 10^{-2} $&$   0.25 \times 10^{-3} $&$   0.53 \times 10^{-2} $&$   0.86 \times 10^{-2}$ \\
  88 &$ 195.6242 $&$  44.0521 $&$   0.2952 $&$   3.8369 $&$   0.92 \times 10^{-3} $&$   0.19 \times 10^{-3} $&$   0.11 \times 10^{-3} $&$   0.80 \times 10^{-2}$ \\
  89 &$ 195.7735 $&$ -24.3063 $&$   0.2030 $&$   6.2071 $&$   0.25 \times 10^{-2} $&$   0.34 \times 10^{-3} $&$  -0.55 \times 10^{-2} $&$   0.92 \times 10^{-2}$ \\
  90 &$ 205.9614 $&$ -39.4838 $&$   0.4430 $&$   4.0854 $&$   0.21 \times 10^{-2} $&$   0.23 \times 10^{-3} $&$   0.25 \times 10^{-1} $&$   0.85 \times 10^{-2}$ \\
  91 &$ 209.5639 $&$ -36.4936 $&$   0.0326 $&$  25.2364 $&$   0.90 \times 10^{-2} $&$   0.12 \times 10^{-2} $&$  -0.29 \times 10^{-1} $&$   0.15 \times 10^{-1}$ \\
  92 &$ 216.6243 $&$  47.0225 $&$   0.3826 $&$   3.9972 $&$   0.12 \times 10^{-2} $&$   0.20 \times 10^{-3} $&$   0.59 \times 10^{-2} $&$   0.83 \times 10^{-2}$ \\
  93 &$ 218.8563 $&$  35.5065 $&$   0.1751 $&$   6.3049 $&$   0.15 \times 10^{-2} $&$   0.28 \times 10^{-3} $&$   0.13 \times 10^{-1} $&$   0.94 \times 10^{-2}$ \\
  94 &$ 226.1795 $&$ -21.9123 $&$   0.0989 $&$   9.1053 $&$   0.26 \times 10^{-2} $&$   0.39 \times 10^{-3} $&$   0.55 \times 10^{-2} $&$   0.11 \times 10^{-1}$ \\
  95 &$ 226.2475 $&$  76.7657 $&$   0.1427 $&$   7.9934 $&$   0.32 \times 10^{-2} $&$   0.35 \times 10^{-3} $&$   0.19 \times 10^{-1} $&$   0.10 \times 10^{-1}$ \\
  96 &$ 228.1552 $&$  75.1923 $&$   0.5450 $&$   3.1858 $&$   0.88 \times 10^{-3} $&$   0.18 \times 10^{-3} $&$   0.29 \times 10^{-2} $&$   0.78 \times 10^{-2}$ \\
  97 &$ 228.4974 $&$  53.1289 $&$   0.1434 $&$   6.7173 $&$   0.12 \times 10^{-2} $&$   0.28 \times 10^{-3} $&$  -0.20 \times 10^{-1} $&$   0.96 \times 10^{-2}$ \\
  98 &$ 229.2177 $&$ -17.2471 $&$   0.1710 $&$   5.6620 $&$   0.17 \times 10^{-2} $&$   0.26 \times 10^{-3} $&$   0.87 \times 10^{-2} $&$   0.93 \times 10^{-2}$ \\
  99 &$ 229.6413 $&$  77.9642 $&$   0.2690 $&$   4.2806 $&$   0.15 \times 10^{-2} $&$   0.23 \times 10^{-3} $&$  -0.54 \times 10^{-2} $&$   0.83 \times 10^{-2}$ \\
 100 &$ 229.9437 $&$  15.2954 $&$   0.0704 $&$  13.4113 $&$   0.64 \times 10^{-2} $&$   0.58 \times 10^{-3} $&$   0.22 \times 10^{-1} $&$   0.13 \times 10^{-1}$ \\
 101 &$ 234.5921 $&$  73.0189 $&$   0.0214 $&$  33.9178 $&$   0.81 \times 10^{-2} $&$   0.16 \times 10^{-2} $&$   0.97 \times 10^{-2} $&$   0.19 \times 10^{-1}$ \\
 102 &$ 236.9552 $&$ -26.6707 $&$   0.1483 $&$   7.1602 $&$   0.17 \times 10^{-2} $&$   0.29 \times 10^{-3} $&$   0.83 \times 10^{-2} $&$   0.10 \times 10^{-1}$ \\
 103 &$ 239.2841 $&$  24.7690 $&$   0.0542 $&$  18.1124 $&$   0.18 \times 10^{-1} $&$   0.68 \times 10^{-3} $&$   0.11 \times 10^{-1} $&$   0.15 \times 10^{-1}$ \\
 104 &$ 239.2893 $&$ -25.9962 $&$   0.4070 $&$   3.4432 $&$   0.14 \times 10^{-2} $&$   0.32 \times 10^{-3} $&$   0.46 \times 10^{-2} $&$   0.74 \times 10^{-2}$ \\
 105 &$ 241.7429 $&$ -30.8853 $&$   0.2708 $&$   4.4853 $&$   0.16 \times 10^{-2} $&$   0.20 \times 10^{-3} $&$   0.65 \times 10^{-2} $&$   0.85 \times 10^{-2}$ \\
 106 &$ 241.7782 $&$ -24.0010 $&$   0.1392 $&$   7.5582 $&$   0.21 \times 10^{-2} $&$   0.28 \times 10^{-3} $&$   0.22 \times 10^{-1} $&$   0.10 \times 10^{-1}$ \\
 107 &$ 241.8563 $&$  51.5311 $&$   0.0700 $&$   9.7408 $&$   0.13 \times 10^{-2} $&$   0.37 \times 10^{-3} $&$  -0.12 \times 10^{-1} $&$   0.12 \times 10^{-1}$ \\
 108 &$ 241.9745 $&$  14.8562 $&$   0.1687 $&$   6.2748 $&$   0.23 \times 10^{-2} $&$   0.26 \times 10^{-3} $&$   0.13 \times 10^{-1} $&$   0.94 \times 10^{-2}$ \\
 109 &$ 243.5701 $&$  67.7603 $&$   0.0834 $&$  10.4394 $&$   0.34 \times 10^{-2} $&$   0.40 \times 10^{-3} $&$   0.61 \times 10^{-2} $&$   0.12 \times 10^{-1}$ \\
 110 &$ 244.3430 $&$ -32.1379 $&$   0.2839 $&$   4.9752 $&$   0.16 \times 10^{-2} $&$   0.21 \times 10^{-3} $&$   0.22 \times 10^{-2} $&$   0.87 \times 10^{-2}$ \\
 111 &$ 244.6968 $&$  32.4925 $&$   0.1535 $&$   6.2505 $&$   0.16 \times 10^{-2} $&$   0.25 \times 10^{-3} $&$  -0.32 \times 10^{-2} $&$   0.93 \times 10^{-2}$ \\
 112 &$ 246.5206 $&$ -26.0574 $&$   0.0468 $&$  15.4600 $&$   0.29 \times 10^{-2} $&$   0.54 \times 10^{-3} $&$   0.19 \times 10^{-1} $&$   0.14 \times 10^{-1}$ \\
 113 &$ 247.1750 $&$ -23.3277 $&$   0.1520 $&$   6.3719 $&$   0.11 \times 10^{-2} $&$   0.23 \times 10^{-3} $&$  -0.37 \times 10^{-2} $&$   0.11 \times 10^{-1}$ \\
 114 &$ 249.8766 $&$ -39.8654 $&$   0.1501 $&$   6.2807 $&$   0.10 \times 10^{-2} $&$   0.22 \times 10^{-3} $&$  -0.58 \times 10^{-2} $&$   0.96 \times 10^{-2}$ \\
 115 &$ 250.9065 $&$ -36.2550 $&$   0.2000 $&$   5.5837 $&$   0.16 \times 10^{-2} $&$   0.21 \times 10^{-3} $&$   0.10 \times 10^{-1} $&$   0.92 \times 10^{-2}$ \\
 116 &$ 252.9668 $&$ -56.0549 $&$   0.0752 $&$  13.1950 $&$   0.26 \times 10^{-2} $&$   0.39 \times 10^{-3} $&$   0.71 \times 10^{-2} $&$   0.13 \times 10^{-1}$ \\
 117 &$ 253.4796 $&$ -33.7242 $&$   0.1913 $&$   5.3897 $&$   0.12 \times 10^{-2} $&$   0.19 \times 10^{-3} $&$  -0.13 \times 10^{-1} $&$   0.94 \times 10^{-2}$ \\
 118 &$ 256.4510 $&$ -65.7122 $&$   0.2195 $&$   5.3944 $&$   0.16 \times 10^{-2} $&$   0.22 \times 10^{-3} $&$   0.33 \times 10^{-2} $&$   0.94 \times 10^{-2}$ \\
 119 &$ 257.3436 $&$ -22.1833 $&$   0.2026 $&$   5.1414 $&$   0.11 \times 10^{-2} $&$   0.16 \times 10^{-3} $&$  -0.24 \times 10^{-2} $&$   0.88 \times 10^{-2}$ \\
 120 &$ 260.0328 $&$ -63.4448 $&$   0.2836 $&$   4.6171 $&$   0.13 \times 10^{-2} $&$   0.19 \times 10^{-3} $&$   0.43 \times 10^{-2} $&$   0.87 \times 10^{-2}$ \\

\hline
\end{tabular}
\end{centering}
\end{table}

\addtocounter{table}{-1}
\begin{table}
\begin{centering}
\caption{\wmap\ observations of \planck\ ESZ clusters}
\label{tab3}
\begin{tabular}{@{}lrrrrrrrr}
\hline
Index & Gal lon ($l$) & Gal lat ($b$) & Redshift & $\theta_{500}$ & $Y^{\rm planck}_{500}$ & $\delta Y^{\rm planck}_{500}$ & $Y^{\rm wmap}_{500}$ & $\delta Y^{\rm wmap}_{500}$ \\
 & (deg) & (deg) & & (arcmin) & (arcmin$^2$) & (arcmin$^2$) & (arcmin$^2$) & (arcmin$^2$) \\
\hline
 121 &$ 262.2535 $&$ -35.3691 $&$   0.2952 $&$   4.8608 $&$   0.19 \times 10^{-2} $&$   0.14 \times 10^{-3} $&$   0.46 \times 10^{-2} $&$   0.88 \times 10^{-2}$ \\
 122 &$ 262.7108 $&$ -40.9137 $&$   0.3900 $&$   3.7414 $&$   0.11 \times 10^{-2} $&$   0.13 \times 10^{-3} $&$  -0.31 \times 10^{-2} $&$   0.79 \times 10^{-2}$ \\
 123 &$ 263.1612 $&$ -23.4141 $&$   0.2266 $&$   5.7940 $&$   0.18 \times 10^{-2} $&$   0.18 \times 10^{-3} $&$   0.27 \times 10^{-2} $&$   0.93 \times 10^{-2}$ \\
 124 &$ 263.2091 $&$ -25.2105 $&$   0.0506 $&$  15.3956 $&$   0.41 \times 10^{-2} $&$   0.48 \times 10^{-3} $&$   0.15 \times 10^{-1} $&$   0.14 \times 10^{-1}$ \\
 125 &$ 263.6653 $&$ -22.5362 $&$   0.1644 $&$   7.1784 $&$   0.36 \times 10^{-2} $&$   0.22 \times 10^{-3} $&$   0.51 \times 10^{-2} $&$   0.99 \times 10^{-2}$ \\
 126 &$ 265.0068 $&$ -48.9483 $&$   0.0590 $&$  15.5042 $&$   0.65 \times 10^{-2} $&$   0.53 \times 10^{-3} $&$   0.15 \times 10^{-3} $&$   0.13 \times 10^{-1}$ \\
 127 &$ 266.0384 $&$ -21.2523 $&$   0.2965 $&$   5.3920 $&$   0.37 \times 10^{-2} $&$   0.18 \times 10^{-3} $&$   0.11 \times 10^{-1} $&$   0.92 \times 10^{-2}$ \\
 128 &$ 266.8423 $&$  25.0788 $&$   0.2542 $&$   5.4913 $&$   0.15 \times 10^{-2} $&$   0.23 \times 10^{-3} $&$   0.27 \times 10^{-2} $&$   0.90 \times 10^{-2}$ \\
 129 &$ 269.3112 $&$ -49.8784 $&$   0.0853 $&$   9.4591 $&$   0.22 \times 10^{-2} $&$   0.36 \times 10^{-3} $&$  -0.98 \times 10^{-2} $&$   0.12 \times 10^{-1}$ \\
 130 &$ 269.5163 $&$  26.4247 $&$   0.0126 $&$  43.3013 $&$   0.12 \times 10^{-1} $&$   0.16 \times 10^{-2} $&$  -0.39 \times 10^{-2} $&$   0.22 \times 10^{-1}$ \\
 131 &$ 271.1969 $&$ -30.9695 $&$   0.3700 $&$   4.0128 $&$   0.11 \times 10^{-2} $&$   0.12 \times 10^{-3} $&$   0.10 \times 10^{-2} $&$   0.80 \times 10^{-2}$ \\
 132 &$ 271.5017 $&$ -56.5590 $&$   0.3000 $&$   4.0513 $&$   0.14 \times 10^{-2} $&$   0.39 \times 10^{-3} $&$  -0.35 \times 10^{-2} $&$   0.82 \times 10^{-2}$ \\
 133 &$ 272.1077 $&$ -40.1502 $&$   0.0589 $&$  16.8371 $&$   0.16 \times 10^{-1} $&$   0.65 \times 10^{-3} $&$   0.10 \times 10^{-1} $&$   0.14 \times 10^{-1}$ \\
 134 &$ 273.6439 $&$  63.2815 $&$   0.1339 $&$   7.5915 $&$   0.28 \times 10^{-2} $&$   0.40 \times 10^{-3} $&$   0.88 \times 10^{-2} $&$   0.10 \times 10^{-1}$ \\
 135 &$ 275.2199 $&$  43.9212 $&$   0.1068 $&$   8.6906 $&$   0.24 \times 10^{-2} $&$   0.46 \times 10^{-3} $&$   0.22 \times 10^{-1} $&$   0.11 \times 10^{-1}$ \\
 136 &$ 277.7522 $&$ -51.7307 $&$   0.4400 $&$   3.5019 $&$   0.15 \times 10^{-2} $&$   0.18 \times 10^{-3} $&$   0.14 \times 10^{-1} $&$   0.77 \times 10^{-2}$ \\
 137 &$ 278.6061 $&$  39.1716 $&$   0.3075 $&$   4.6768 $&$   0.19 \times 10^{-2} $&$   0.30 \times 10^{-3} $&$   0.13 \times 10^{-1} $&$   0.86 \times 10^{-2}$ \\
 138 &$ 280.1968 $&$  47.8167 $&$   0.1557 $&$   6.1951 $&$   0.23 \times 10^{-2} $&$   0.34 \times 10^{-3} $&$  -0.17 \times 10^{-4} $&$   0.99 \times 10^{-2}$ \\
 139 &$ 282.4933 $&$  65.1743 $&$   0.0766 $&$  12.0485 $&$   0.53 \times 10^{-2} $&$   0.69 \times 10^{-3} $&$   0.34 \times 10^{-1} $&$   0.13 \times 10^{-1}$ \\
 140 &$ 284.4608 $&$  52.4371 $&$   0.4414 $&$   3.6666 $&$   0.16 \times 10^{-2} $&$   0.25 \times 10^{-3} $&$   0.47 \times 10^{-2} $&$   0.78 \times 10^{-2}$ \\
 141 &$ 284.9936 $&$ -23.7081 $&$   0.3900 $&$   3.9007 $&$   0.13 \times 10^{-2} $&$   0.12 \times 10^{-3} $&$   0.89 \times 10^{-2} $&$   0.81 \times 10^{-2}$ \\
 142 &$ 285.6352 $&$ -17.2466 $&$   0.3500 $&$   3.5369 $&$   0.90 \times 10^{-3} $&$   0.15 \times 10^{-3} $&$   0.53 \times 10^{-2} $&$   0.77 \times 10^{-2}$ \\
 143 &$ 286.5869 $&$ -31.2510 $&$   0.2100 $&$   5.6231 $&$   0.14 \times 10^{-2} $&$   0.21 \times 10^{-3} $&$  -0.34 \times 10^{-2} $&$   0.91 \times 10^{-2}$ \\
 144 &$ 286.9927 $&$  32.9157 $&$   0.3900 $&$   5.0661 $&$   0.34 \times 10^{-2} $&$   0.33 \times 10^{-3} $&$   0.20 \times 10^{-1} $&$   0.91 \times 10^{-2}$ \\
 145 &$ 288.6160 $&$ -37.6562 $&$   0.1270 $&$   7.1712 $&$   0.29 \times 10^{-2} $&$   0.31 \times 10^{-3} $&$   0.90 \times 10^{-3} $&$   0.98 \times 10^{-2}$ \\
 146 &$ 292.5194 $&$  21.9886 $&$   0.3000 $&$   5.1319 $&$   0.21 \times 10^{-2} $&$   0.34 \times 10^{-3} $&$   0.18 \times 10^{-1} $&$   0.91 \times 10^{-2}$ \\
 147 &$ 294.6674 $&$ -37.0299 $&$   0.2742 $&$   4.4588 $&$   0.15 \times 10^{-2} $&$   0.20 \times 10^{-3} $&$   0.68 \times 10^{-2} $&$   0.84 \times 10^{-2}$ \\
 148 &$ 295.3328 $&$  23.3359 $&$   0.1190 $&$   7.2625 $&$   0.23 \times 10^{-2} $&$   0.43 \times 10^{-3} $&$   0.15 \times 10^{-2} $&$   0.10 \times 10^{-1}$ \\
 149 &$ 296.4139 $&$ -32.4851 $&$   0.0613 $&$  12.4130 $&$   0.25 \times 10^{-2} $&$   0.40 \times 10^{-3} $&$   0.19 \times 10^{-1} $&$   0.13 \times 10^{-1}$ \\
 150 &$ 303.7589 $&$  33.6561 $&$   0.0544 $&$  13.6334 $&$   0.47 \times 10^{-2} $&$   0.80 \times 10^{-3} $&$   0.14 \times 10^{-1} $&$   0.13 \times 10^{-1}$ \\
 151 &$ 304.4970 $&$  32.4428 $&$   0.0554 $&$  13.9820 $&$   0.37 \times 10^{-2} $&$   0.82 \times 10^{-3} $&$   0.49 \times 10^{-2} $&$   0.13 \times 10^{-1}$ \\
 152 &$ 304.6716 $&$ -31.6688 $&$   0.1934 $&$   5.1350 $&$   0.11 \times 10^{-2} $&$   0.20 \times 10^{-3} $&$  -0.14 \times 10^{-2} $&$   0.89 \times 10^{-2}$ \\
 153 &$ 304.8952 $&$  45.4509 $&$   0.0473 $&$  17.6562 $&$   0.84 \times 10^{-2} $&$   0.10 \times 10^{-2} $&$  -0.25 \times 10^{-1} $&$   0.14 \times 10^{-1}$ \\
 154 &$ 306.6839 $&$  61.0626 $&$   0.0845 $&$  11.5603 $&$   0.52 \times 10^{-2} $&$   0.65 \times 10^{-3} $&$  -0.15 \times 10^{-2} $&$   0.12 \times 10^{-1}$ \\
 155 &$ 306.8008 $&$  58.6075 $&$   0.0845 $&$  11.8327 $&$   0.43 \times 10^{-2} $&$   0.68 \times 10^{-3} $&$  -0.43 \times 10^{-2} $&$   0.13 \times 10^{-1}$ \\
 156 &$ 311.9973 $&$  30.7170 $&$   0.0480 $&$  19.4676 $&$   0.12 \times 10^{-1} $&$   0.13 \times 10^{-2} $&$   0.15 \times 10^{-1} $&$   0.14 \times 10^{-1}$ \\
 157 &$ 313.3604 $&$  61.1166 $&$   0.1832 $&$   7.4682 $&$   0.39 \times 10^{-2} $&$   0.44 \times 10^{-3} $&$   0.78 \times 10^{-2} $&$   0.95 \times 10^{-2}$ \\
 158 &$ 313.8723 $&$ -17.1066 $&$   0.1530 $&$   8.1257 $&$   0.43 \times 10^{-2} $&$   0.38 \times 10^{-3} $&$   0.16 \times 10^{-1} $&$   0.11 \times 10^{-1}$ \\
 159 &$ 315.7076 $&$ -18.0430 $&$   0.1050 $&$   9.6202 $&$   0.49 \times 10^{-2} $&$   0.46 \times 10^{-3} $&$   0.16 \times 10^{-1} $&$   0.11 \times 10^{-1}$ \\
 160 &$ 316.3467 $&$  28.5433 $&$   0.0391 $&$  24.8271 $&$   0.21 \times 10^{-1} $&$   0.17 \times 10^{-2} $&$   0.17 \times 10^{-1} $&$   0.16 \times 10^{-1}$ \\
 161 &$ 318.1334 $&$ -29.5778 $&$   0.2170 $&$   5.1100 $&$   0.17 \times 10^{-2} $&$   0.26 \times 10^{-3} $&$   0.31 \times 10^{-2} $&$   0.86 \times 10^{-2}$ \\
 162 &$ 321.9628 $&$ -47.9754 $&$   0.0940 $&$   9.9841 $&$   0.29 \times 10^{-2} $&$   0.32 \times 10^{-3} $&$   0.13 \times 10^{-1} $&$   0.12 \times 10^{-1}$ \\
 163 &$ 324.4983 $&$ -44.9709 $&$   0.0951 $&$   8.9942 $&$   0.21 \times 10^{-2} $&$   0.30 \times 10^{-3} $&$  -0.13 \times 10^{-1} $&$   0.11 \times 10^{-1}$ \\
 164 &$ 332.2345 $&$ -46.3695 $&$   0.0980 $&$  10.4829 $&$   0.47 \times 10^{-2} $&$   0.38 \times 10^{-3} $&$   0.18 \times 10^{-2} $&$   0.12 \times 10^{-1}$ \\
 165 &$ 332.8879 $&$ -19.2803 $&$   0.1470 $&$   7.1985 $&$   0.24 \times 10^{-2} $&$   0.32 \times 10^{-3} $&$   0.59 \times 10^{-2} $&$   0.97 \times 10^{-2}$ \\
 166 &$ 335.5923 $&$ -46.4637 $&$   0.0760 $&$  11.4352 $&$   0.46 \times 10^{-2} $&$   0.42 \times 10^{-3} $&$   0.19 \times 10^{-1} $&$   0.13 \times 10^{-1}$ \\
 167 &$ 336.5914 $&$ -55.4487 $&$   0.0965 $&$   9.5308 $&$   0.32 \times 10^{-2} $&$   0.32 \times 10^{-3} $&$   0.12 \times 10^{-1} $&$   0.12 \times 10^{-1}$ \\
 168 &$ 340.8860 $&$ -33.3489 $&$   0.0556 $&$  18.6213 $&$   0.15 \times 10^{-1} $&$   0.78 \times 10^{-3} $&$   0.18 \times 10^{-1} $&$   0.15 \times 10^{-1}$ \\
 169 &$ 340.9585 $&$  35.1160 $&$   0.2357 $&$   6.0862 $&$   0.17 \times 10^{-2} $&$   0.38 \times 10^{-3} $&$  -0.14 \times 10^{-1} $&$   0.98 \times 10^{-2}$ \\
 170 &$ 342.3172 $&$ -34.9065 $&$   0.2320 $&$   4.8257 $&$   0.16 \times 10^{-2} $&$   0.24 \times 10^{-3} $&$  -0.41 \times 10^{-2} $&$   0.91 \times 10^{-2}$ \\
 171 &$ 342.8155 $&$ -30.4605 $&$   0.0600 $&$  10.9597 $&$   0.24 \times 10^{-2} $&$   0.51 \times 10^{-3} $&$   0.39 \times 10^{-1} $&$   0.13 \times 10^{-1}$ \\
 172 &$ 345.4068 $&$ -39.3440 $&$   0.0448 $&$  23.7182 $&$   0.60 \times 10^{-2} $&$   0.18 \times 10^{-2} $&$  -0.16 \times 10^{-1} $&$   0.15 \times 10^{-1}$ \\
 173 &$ 346.5981 $&$  35.0477 $&$   0.2226 $&$   5.3936 $&$   0.26 \times 10^{-2} $&$   0.31 \times 10^{-3} $&$   0.31 \times 10^{-3} $&$   0.89 \times 10^{-2}$ \\
 174 &$ 347.1876 $&$ -27.3538 $&$   0.2371 $&$   4.9553 $&$   0.12 \times 10^{-2} $&$   0.24 \times 10^{-3} $&$   0.25 \times 10^{-2} $&$   0.93 \times 10^{-2}$ \\
 175 &$ 349.4626 $&$ -59.9475 $&$   0.3475 $&$   4.9888 $&$   0.25 \times 10^{-2} $&$   0.19 \times 10^{-3} $&$   0.15 \times 10^{-1} $&$   0.92 \times 10^{-2}$ \\

\hline
\end{tabular}
\end{centering}
\end{table}

\end{appendix}


\begin{thebibliography}{99}
\bibitem[\protect\citeauthoryear{Aghanim et al.}{1997}]{Aghanim97}
Aghanim N., de Luca A., Bouchet F.~R., Gispert R., \& Puget J.~L.,
1997, A\&A, 325, 9

\bibitem[\protect\citeauthoryear{Akritas \& Bershady}{1996}]{Akritas96} Akritas M. G., \& Bershady M.
A., 1996, ApJ, 470, 706

\bibitem[\protect\citeauthoryear{Arnaud et al.}{2010}]{Arnaud10}
Arnaud M. et al., 2010, A\&A, 517, 92

\bibitem[\protect\citeauthoryear{Birkinshaw \& Gull}{1978}]{Birkinshaw78}
Birkinshaw M., \& Gull S.~F., 1978, Nature, 274, 111

\bibitem[\protect\citeauthoryear{Birkinshaw}{1999}]{Birkinshaw99}
Birkinshaw M., 1999, Phys. Rep., 310, 97

\bibitem[\protect\citeauthoryear{Carlstrom et al.}{2002}]{Carlstrom02}
Carlstrom J.~E., Holder G.~P., \& Reese E.~D., 2002, ARA\&A, 40,
643

\bibitem[\protect\citeauthoryear{Carlstrom et al.}{2011}]{Carlstrom11}
Carlstrom J.~E., 2011, Publ. Astron. Soc. Pac. 123, 568-581

\bibitem[\protect\citeauthoryear{Clutton-Brock}{1967}]{Clutton-Brock67}
Clutton-Brock M., 1967, Technometrics, 9, 261

\bibitem[\protect\citeauthoryear{Douspis, Aghanim \& Langer}{2006}]{Douspis06}
Douspis M., Aghanim N., \& Langer M., 2006, A\&A, 456, 819

\bibitem[\protect\citeauthoryear{Gorski et al.}{2005}]{Gorski05}
Gorski K. M., Hivon E., Banday A. J., Wandelt B. D., Hansen F. K.,
Reinecke M.,\& Bartelmann M., 2005, ApJ 622, 759


\bibitem[\protect\citeauthoryear{Haiman, Mohr \& Holder}{2001}]{Haiman01}
Haiman Z., Mohr J.~J., \& Holder G.~P., 2001, ApJ, 553, 545


\bibitem[\protect\citeauthoryear{Hinshaw et al.}{2007}]{Hinshaw07}
Hinshaw G. et al., 2007, ApJS, 170, 288

\bibitem[\protect\citeauthoryear{Komatsu et al.}{2011}]{Komatsu11}
Komatsu E., 2011, ApJS., 192, 18

\bibitem[\protect\citeauthoryear{LAMBDA website}{}]{Lambdaweb}
WMAP Lambda website: http://lambda.gsfc.nasa.gov/

\bibitem[\protect\citeauthoryear{Levine, Schulz, \& White}{2002}]{Levine02}
Levine E.~S., Schulz A.~E., \& White M., 2002, ApJ, 577, 569

\bibitem[\protect\citeauthoryear{Majumdar \& Mohr}{2004}]{Majumdar04}
Majumdar S., \& Mohr J.~J., 2004, ApJ, 613, 41

\bibitem[\protect\citeauthoryear{Marriage et al.}{2011}]{Marriage11}
Marriage T. A., 2011, ApJ, 737, 61

\bibitem[\protect\citeauthoryear{Melin et al.}{2006}]{Melin06}
Melin J., Bartlett J.~G., Delabrouille J., 2006, A\&A, 459, 341


\bibitem[\protect\citeauthoryear{McCarthy et al.}{2007}]{McCarthy07}
McCarthy I.~G., Bower R.~G., Balogh M.~L., 2007, MNRAS, 377,
1457-1463

\bibitem[\protect\citeauthoryear{Nagai, Kravtsov \& Vikhlinin}{2007}]{Nagai07}
Nagai D., Kravtsov A. V., \& Vikhlinin A., 2007, ApJ, 668, 1


\bibitem[\protect\citeauthoryear{Planck Collaboration}{2005}]{Planck05}
The Planck Collaboration, \textit{Planck: The Scientific
Programme}, European Space Agency Vol. No. ESA¨CSCI (2005)1,
edited by G. Efstathiou, et al. (ESA Publications, Noordwijk,
Netherlands, 2005).

\bibitem[\protect\citeauthoryear{Planck Collaboration VII}{2011}]{Planck11b}
Planck Collaboration VII, 2011, A\&A, 536, 7

\bibitem[\protect\citeauthoryear{Planck Collaboration VIII}{2011}]{Planck11a}
Planck Collaboration VIII, 2011, A\&A, 536, 8

\bibitem[\protect\citeauthoryear{Planck Collaboration IX}{2011}]{Planck11c}
Planck Collaboration IX, 2011, A\&A, 536, 9

\bibitem[\protect\citeauthoryear{Planck Collaboration XII}{2011}]{Planck11d}
Planck Collaboration XII, 2011, A\&A, 536, 12

\bibitem[\protect\citeauthoryear{Press et al.}{2007}]{Press07}
Press W. H., Teukolsky S. A., Vetterling W. T., \& Flannery B. P.,
\textit{Numerical Receipes} 3rd edition, 2007, Cambridge
University Press.


\bibitem[\protect\citeauthoryear{Tegmark \& Efstathiou}{1996}]{Tegmark96}
Tegmark M. \& Efstathiou G., 1996, MNRAS, 281, 1297

\bibitem[\protect\citeauthoryear{Tegmark \& de Oliveira-Costa}{1998}]{Tegmark98}
Tegmark M. \& de Oliveira-Costa A., 1998, ApJ, 500, 83

\bibitem[\protect\citeauthoryear{Shaw, Rudd \& Nagai}{2012}]{Shaw11}
Shaw L. D., Rudd D. H., \& Nagai D., 2012, ApJ, 756, 15

\bibitem[\protect\citeauthoryear{Sunyaev \& Zeldovich}{1972}]{Sunyaev72}
Sunyaev R.~A., \& Zeldovich Y.~B., 1972, Comments on Astrophysics
and Space Physics, 4, 173

\bibitem[\protect\citeauthoryear{Sunyaev \& Zeldovich}{1980}]{Sunyaev80}
Sunyaev R.~A., \& Zeldovich I.~B., 1980, ARA\&A, 18, 537

\bibitem[\protect\citeauthoryear{Weller, Battye \& Kneissl}{2002}]{Weller02}
Weller J., Battye R.~A., \& Kneissl R., 2002, Phys. Rev. Lett.,
88, 231301

\bibitem[\protect\citeauthoryear{Whitbourn, Shanks \& Sawangwit}{2011}]{Witbourn11}
Whitbourn J. R., Shanks T., \& Sawangwit U., arXiv:1107.2654
[astro-ph].

\bibitem[\protect\citeauthoryear{Zwart et al.}{2008}]{Zwart08}
Zwart J.~T.~L. et~al., 2008, MNRAS, 391, 1545

\end{thebibliography}
\end{document}